\documentclass[aps,prb,twocolumn,superscriptaddress,showpacs]{revtex4}

\usepackage{amsmath,amssymb}
\usepackage{graphicx}
\usepackage[hypertex,breaklinks=true,colorlinks=false]{hyperref}

\def\rnum#1{\expandafter{%
\romannumeral #1}}
\def\Rnum#1{\uppercase\expandafter{%
\romannumeral #1}}

\newcommand{\bol}[1]{\boldsymbol #1}

\newcommand{\PRL}[3]{Phys. Rev. Lett. {\bf #1},
\href{http://link.aps.org/abstract/PRL/v#1/e#2}{#2} (#3)}
\newcommand{\PRLp}[3]{Phys. Rev. Lett. {\bf #1},
\href{http://link.aps.org/abstract/PRL/v#1/p#2}{#2} (#3)}

\newcommand{\PRB}[3]{Phys. Rev. B {\bf #1},
\href{http://link.aps.org/abstract/PRB/v#1/e#2}{#2} (#3)}
\newcommand{\PRBp}[3]{Phys. Rev. B {\bf #1},
\href{http://link.aps.org/abstract/PRB/v#1/p#2}{#2} (#3)}

\newcommand{\PRBRp}[3]{Phys. Rev. B {\bf #1},
\href{http://link.aps.org/abstract/PRB/v#1/e#2}{R#2} (#3)}

\newcommand{\condmat}[1]{arXiv:cond-mat/\href{http://arxiv.org/abs/cond-mat/#1}{#1}}

\begin{document}


\title{Coefficients of bosonized dimer operators in 
spin-$\frac{\bol 1}{\bol 2}$ XXZ chains and their applications}


\author{Shintaro Takayoshi}
\affiliation{Institute for Solid State Physics, University of Tokyo, 
Kashiwa, Chiba 277-8581, Japan}
\author{Masahiro Sato}
\affiliation{Condensed Matter Theory Laboratory, RIKEN, Wako, 
Saitama 351-0198, Japan}
\affiliation{Department of Physics and Mathematics, 
Aoyama-Gakuin University, Sagamihara, 
Kanagawa 229-8558, Japan}


\date{\today}

\begin{abstract}
Comparing numerically evaluated excitation gaps of 
dimerized spin-$\frac{1}{2}$ XXZ chains with the gap formula for 
the 
low-energy effective sine-Gordon theory, we 
determine coefficients 
$d_{xy}$ and $d_{z}$ of bosonized dimerization operators in 
spin-$\frac{1}{2}$ XXZ chains, 
which are defined as $(-1)^{j}(S_{j}^xS_{j+1}^x+S_{j}^yS_{j+1}^y)=
d_{xy}\sin(\sqrt{4\pi}\phi(x))+\cdots$ and 
$(-1)^{j}S_{j}^{z}S_{j+1}^{z}=d_{z}\sin(\sqrt{4\pi}\phi(x))+\cdots$. 
We also calculate 
the coefficients of both spin and dimer operators for 
the spin-$\frac{1}{2}$ Heisenberg antiferromagnetic 
chain with a nearest-neighbor coupling $J$ and 
a next-nearest-neighbor coupling $J_2=0.2411J$. 
As applications of these coefficients, we present 
ground-state phase diagrams of dimerized spin chains 
in a magnetic field and antiferromagnetic spin ladders 
with a four-spin interaction. 
The optical conductivity and electric polarization of one-dimensional 
Mott insulators with Peierls instability are also evaluated 
quantitatively. 
\end{abstract}



\maketitle


\section{Introduction}
Quantum magnets in one dimension are a basic class of many-body systems 
in condensed matter and statistical physics 
(see e.g., Refs.~\onlinecite{Giamarchi,Affleck}). 
They have offered various kinds of topics 
in both experimental and theoretical studies for a long time. 
In particular, the spin-$\frac{1}{2}$ XXZ chain is a simple though
realistic system in this field. The Hamiltonian is defined by 
\begin{equation}
{\cal H}^{\rm XXZ}=J\sum_{j}(S_{j}^{x}S_{j+1}^{x}+S_{j}^{y}S_{j+1}^{y}
+\Delta_{z}S_{j}^{z}S_{j+1}^{z}),
\label{eq:XXZ}
\end{equation}
where $S_{j}^{\alpha}$ is $\alpha$-component of a 
spin-$\frac{1}{2}$ operator on $j$-th site, 
$J>0$ is the exchange coupling constant, 
and $\Delta_{z}$ is the anisotropy parameter. 
This model 
is exactly solved by integrability methods,~\cite{Korepin,Takahashi} 
and the ground-state phase diagram has been completed. 
Three phases appear depending on $\Delta_{z}$; 
the antiferromagnetic (AF) phase with a N\'eel order 
$\langle S_{j}^{z}\rangle=-\langle S_{j+1}^{z}\rangle$ 
($\Delta_z>1$), the critical Tomonaga-Luttinger liquid (TLL) phase 
($-1<\Delta_z\leq 1$), and the fully polarized phase with 
$\langle S_{j}^{z}\rangle=1/2$ ($\Delta_z\leq -1$). 
In and around the TLL phase, the low-energy and long-distance properties 
can be understood via effective field theory techniques such 
as bosonization and conformal field 
theory (CFT).~\cite{Giamarchi,Affleck,Tsvelik,Gogolin,Francesco} 
These theoretical results nicely explain experiments of several quasi 
one-dimensional (1D) magnets. 
The deep knowledge of this model is also useful for analyzing 
plentiful related magnetic systems, such as 
spin-$\frac{1}{2}$ chains with some perturbations 
(e.g. external fields,~\cite{Alcaraz95} 
additional magnetic anisotropies,~\cite{Oshikawa97,Affleck99,Essler98,Kuzmenko09}
dimerization~\cite{Haldane82,Papenbrock03,Orignac04}), 
coupled spin chains,~\cite{Shelton96,Kim2000} 
spatially anisotropic 2D or 3D spin 
systems,~\cite{Starykh04,Balents07,Starykh07} etc.

A recent direction 
of studying spin chains is to 
establish solid correspondences between the model~(\ref{eq:XXZ}) 
and its effective theory. For example, Lukyanov and 
his collaborators~\cite{Lukyanov97,Lukyanov99,Lukyanov03} 
have analytically predicted 
coefficients of bosonized spin operators in the TLL phase. 
Hikihara and Furusaki~\cite{Hikihara98,Hikihara04} 
have also determined them numerically in the same chains 
with and without a uniform Zeeman term. 
Using these results, one can now calculate amplitudes of 
spin correlation functions as well as their critical exponents. 
Furthermore, effects of perturbations on an XXZ chain can also be 
calculated with high accuracy. 
It therefore becomes possible to quantitatively compare 
theoretical and experimental results in quasi 1D magnets. 
The purpose of the present study is to attach a new relationship 
between the spin-$\frac{1}{2}$ XXZ chain and its bosonized effective theory. 
Namely, we numerically evaluate coefficients of 
bosonized dimer operators in the TLL phase of the XXZ chain. 
Dimer operators $(-1)^jS_j^\alpha S_{j+1}^\alpha$, 
as well as spin operators, are 
fundamental degrees of freedom in spin-$\frac{1}{2}$ AF chains. 
In fact, the leading terms of both bosonized spin and dimer 
operators have the same scaling dimension $1/2$  
at the $SU(2)$-symmetric AF point $\Delta_z=1$ (see Sec.~\ref{sec:dimer}).

In Refs.~\onlinecite{Hikihara98,Hikihara04}, Hikihara and Furusaki have 
used density-matrix renormalization-group (DMRG) method 
in an efficient manner 
in order to accurately evaluate coefficients of spin operators 
of an XXZ chain in a magnetic field. 
Instead of such a direct powerful method, we 
utilize the relationship between a dimerized XXZ chain and its effective 
sine-Gordon theory~\cite{Essler98,Essler04} to 
determine the coefficients of dimer operators 
(defined in Sec.~\ref{sec:dimer}), i.e., excitation gaps in dimerized spin 
chains are evaluated by numerical diagonalization method and are 
compared with the gap formula of the effective sine-Gordon theory. 
In other words, we derive the information on uniform 
spin-$\frac{1}{2}$ XXZ chains from dimerized (deformed) chains. 
Moreover, we also determine the coefficients of both spin and dimer 
operators for the spin-$\frac{1}{2}$ Heisenberg (i.e., XXX) AF chain with 
an additional next-nearest-neighbor (NNN) coupling $J_2=0.2411J$ 
in the similar strategy. As seen in Sec.~\ref{subsec:SU2_dimer}, 
evaluated coefficients are more reliable for the $J$-$J_2$ model,
since the marginal terms vanish in its effective theory.

The plan of this paper is as follows. 
In Sec.~\ref{sec:dimer}, we shortly summarize the bosonization of 
XXZ spin chains. Both the XXZ chain with dimerization 
and the chain in a staggered magnetic field are mapped to a sine-Gordon model. 
We also consider the AF Heisenberg chain with NNN coupling $J_2=0.2411J$. 
In Sec.~\ref{sec:delta}, we explain how to obtain the coefficients of 
dimer and spin operators by using numerical diagonalization method. 
The evaluated coefficients are listed 
in Tables~\ref{tb:dimer_coeff} and \ref{tb:a1} 
and Fig.~\ref{fig:dimer_coeff}. These are the main results of this paper. 
For comparison, the same dimer coefficients are also calculated 
by using the formula of the ground-state energy of the sine-Gordon model. 
We find that the coefficients fixed by the gap formula are more reliable. 
We apply these coefficients to several systems and physical quantities 
related to an XXZ chain (dimerized spin chains under a magnetic field, 
spin ladders with a four-spin exchange and 
optical response of dimerized 1D Mott insulators) in Sec.~\ref{sec:apply}. 
Finally our results are summarized in Sec.~\ref{sec:con}.

\section{Dimerized chain and sine-Gordon model}
\label{sec:dimer}
In this section, we explain the relationship between a dimerized XXZ 
chain and the corresponding sine-Gordon theory 
in the easy-plane region $-1<\Delta_z\leq 1$.
XXZ chains in a staggered field and the AF Heisenberg chain with 
NNN coupling $J_2=0.2411J$ are also discussed. 
The coefficients of dimer operators are 
defined in Eq.~(\ref{eq:dimer}).

\subsection{Bosonization of spin-$\frac{1}{2}$ XXZ chain}
\label{subsec:XXZ}
We first review the effective theory 
for undimerized spin chain~(\ref{eq:XXZ}). 
According to the standard strategy, 
XXZ Hamiltonian (\ref{eq:XXZ}) is bosonized as
\begin{align}
{\cal H}_{\rm eff}^{\rm XXZ}=\int{\rm d}x\Big\{\frac{v}{2}
[K^{-1}(\partial_{x}\phi)^{2}+K(\partial_{x}\theta)^{2}]  \nonumber\\
-v\frac{\lambda}{2\pi}\cos\big(\sqrt{16\pi}\phi\big)+\cdots\Big\},
\label{eq:XXZ_eff_H}
\end{align}
in the TLL phase. 
Here, $\phi(x)$ and $\theta(x)$ are dual scalar fields, 
which satisfy the commutation relation, 
\begin{equation}
[\phi(x),\theta(x')]=-{\rm i}\vartheta_{\rm step}(x-x'),
\end{equation}
with $x=j a$ ($a$ is the lattice spacing). 
As we see in Eq.~(\ref{eq:normalization}), 
$\cos(\sqrt{16\pi}\phi)$ is irrelevant in $-1<\Delta_z<1$, and becomes 
marginal at the $SU(2)$-symmetric AF Heisenberg point $\Delta_z=1$. 
The coupling constant $\lambda$ has been determined 
exactly.~\cite{Lukyanov98,Lukyanov03} 
Two quantities $K$ and $v$ denote the TLL parameter and spinon velocity, 
respectively, which can be exactly evaluated from 
Bethe ansatz:~\cite{Giamarchi,Cabra98}
\begin{subequations}
\label{eq:K_v}
\begin{align}
K=&\frac{\pi}{2(\pi-\cos^{-1}\Delta_{z})}=\frac{1}{4\pi R^{2}}
=\frac{1}{2\eta},\\
v=&Ja\frac{\pi\sqrt{1-\Delta_{z}^{2}}}{2\cos^{-1}\Delta_{z}}
=Ja\frac{\sin(\pi\eta)}{2(1-\eta)}.
\end{align}
\end{subequations}
Here we have introduced new parameters $\eta$ and $R$. 
The former is the critical exponent of two-point spin correlation functions
and used in the discussion below. 
The latter is called the compactification radius.
It fixes the periodicity 
of fields $\phi$ and $\theta$ as $\phi/\sqrt{K}\sim\phi/\sqrt{K}+2\pi R$ 
and $\sqrt{K}\theta\sim \sqrt{K}\theta+1/R$. 
Using the scalar fields $\phi$ and $\theta$, 
we can obtain the bosonized representation of spin operators: 
\begin{subequations}
\label{eq:spin_boson}
\begin{align}
S_{j}^{z}\approx&\frac{a}{\sqrt{\pi}}\partial_{x}\phi
+(-1)^{j}a_{1}\cos(\sqrt{4\pi}\phi)+\cdots,\\
S_{j}^{+}\approx&{\rm e}^{{\rm i}\sqrt{\pi}\theta}
\left[b_{0}(-1)^{j}+b_{1}\cos(\sqrt{4\pi}\phi)+\cdots\right],
\end{align}
\end{subequations}
where $a_n$ and $b_n$ are non-universal constants, and some of them 
with small $n$ have been determined accurately in 
Refs.~\onlinecite{Lukyanov97,Lukyanov99,Lukyanov03,Hikihara98,Hikihara04}. 
In this formalism, vertex operators are normalized 
as~\cite{Lukyanov97,Lukyanov99,Lukyanov03}
\begin{equation}
\langle{\rm e}^{{\rm i}q\phi(x)}{\rm e}^{-{\rm i}q\phi(x')}\rangle
=\left(\frac{a}{|x-x'|}\right)^{\frac{Kq^{2}}{2\pi}}
{\rm at}\;|x-x'|\gg a.
\label{eq:normalization}
\end{equation}
This means that the operator ${\rm e}^{{\rm i}q\phi(x)}$ has scaling 
dimension $Kq^{2}/(4\pi)$. 

In addition to the spin operators, the bosonized forms of 
the dimer operators are known to 
be~\cite{Giamarchi,Affleck,Tsvelik,Gogolin} 
\begin{subequations}
\label{eq:dimer}
\begin{align}
(-1)^{j}(S_{j}^{x}S_{j+1}^{x}+S_{j}^{y}S_{j+1}^{y})
\approx&d_{xy}\sin(\sqrt{4\pi}\phi)+\cdots,\\
(-1)^{j}S_{j}^{z}S_{j+1}^{z}
\approx&d_{z}\sin(\sqrt{4\pi}\phi)+\cdots.
\end{align}
\end{subequations}
In contrast to the spin operators, the coefficients $d_{xy}$ and $d_z$ 
have never been evaluated so far. To determine them is 
the subject of this paper. It seems to be possible to calculate $d_{xy,z}$ 
by utilizing Eq.~(\ref{eq:spin_boson}) and operator-product-expansion 
(OPE) technique,~\cite{Gogolin,Tsvelik,Francesco} 
but it requires the correct values of all the 
factors $a_n$ and $b_n$.~\cite{Hikihara04} 
Therefore, we should interpret that the 
dimer coefficients $d_{xy,z}$ are independent of spin coefficients 
$a_n$ and $b_n$.


\subsection{Bosonization of dimerized spin chain}
\label{subsec:dimer}
\begin{figure}
\begin{center}
\includegraphics[width=0.35\textwidth]{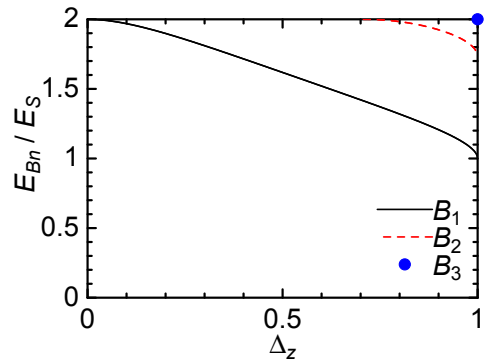}
\end{center}
\caption{(Color online) Ratio of the $n$-th breather mass 
$E_{B_n}$ to the soliton mass $E_S$ as a function of 
the XXZ anisotropy $\Delta_z$ in the sine-Gordon 
model~(\ref{eq:XXZ_dimer_eff}) or (\ref{eq:XXZ_stag_eff}).}
\label{fig:spectrum_SG}
\end{figure}

Next, let us consider a bond-alternating XXZ chain whose Hamiltonian is 
given as
\begin{align}
{\cal H}^{{\rm XXZ}\mathchar`-\delta}=&J\sum_{j}\left[
(1+(-1)^{j}\delta_{xy})(S_{j}^{x}S_{j+1}^{x}
+S_{j}^{y}S_{j+1}^{y})\right.\nonumber\\
&\left.+(\Delta_{z}+(-1)^{j}\delta_{z})
S_{j}^{z}S_{j+1}^{z}\right].
\label{eq:dimer_XXZ}
\end{align}
In the weak dimerization regime of $|\delta_{xy,z}|\ll 1$, 
the bosonization is applicable and the dimerization terms can be treated 
perturbatively. From the formula~(\ref{eq:dimer}), the effective Hamiltonian 
of Eq.~(\ref{eq:dimer_XXZ}) is 
\begin{align}
{\cal H}_{\rm eff}^{{\rm XXZ}\mathchar`-\delta}&=\int{\rm d}x\Big\{
\frac{v}{2}[K^{-1}(\partial_{x}\phi)^{2}+K(\partial_{x}\theta)^{2}] 
\nonumber\\
&+\frac{J}{a}(\delta_{xy}d_{xy}+\delta_{z}d_{z})
\sin(\sqrt{4\pi}\phi)+\cdots \Big\}.
\label{eq:XXZ_dimer_eff}
\end{align}
Here, we have neglected all of the irrelevant terms 
including $\cos(\sqrt{16\pi}\phi)$. 
This is nothing but an integrable sine-Gordon model (see e.g., 
Refs.~\onlinecite{Essler98,Essler04} and references therein). 
The $\sin(\sqrt{4\pi}\phi)$ term has a scaling dimension $K$, 
and is relevant when $K<2$, i.e., $-0.7071<\Delta_z\leq 1$. 
In this case, an excitation gap opens and a dimerization 
$\langle S_{j}^{\alpha}S_{j+1}^{\alpha}
-S_{j+1}^{\alpha}S_{j+2}^{\alpha}\rangle\neq0$ occurs.
The excitation spectrum of the sine-Gordon model 
has been known,~\cite{Essler98,Essler04} and three types of 
elementary particles appear; a soliton, 
the corresponding antisoliton, and bound states 
of the soliton and the antisoliton (called breathers). 
The soliton and antisoliton have the same mass gap $E_S$. 
There exist $[4\eta-1]$ breathers, in which $[A]$ stands for the integer 
part of $A$. The mass of soliton and 
$n$-th breather $E_{B_n}$ are related as follows.
\begin{equation}
E_{B_n}=2 E_S \sin\left(\frac{n\pi}{2(4\eta-1)}\right), 
\quad n=1,\cdots,[4\eta-1]. \label{eq:mass_relation}
\end{equation}
The breather mass in units of the soliton mass 
is shown in Fig.~\ref{fig:spectrum_SG} 
as a function of $\Delta_z$. Note that there is no breather in the 
ferromagnetic side $\Delta_z<0$, and the lightest breather with 
mass $E_{B_1}$ is always heavier than the soliton 
in the present easy-plane regime. 
Following Refs.~\onlinecite{Zamolodchikov95,Lukyanov97}, the soliton mass 
is also analytically represented as 
\begin{align}
\frac{E_S}{J}=&\frac{v}{Ja}\frac{2}{\sqrt{\pi}}
\frac{\Gamma\left(\frac{1}{8\eta-2}\right)}{\Gamma\left(\frac{2}{4-1/\eta}
\right)}
\nonumber\\
&\times\left[\frac{Ja}{v}\frac{\pi (\delta_{xy}d_{xy}+\delta_{z}d_{z})}{2}
\frac{\Gamma\left(\frac{4-1/\eta}{4}\right)}
{\Gamma\left(\frac{1}{4\eta}\right)} \right]^{\frac{2}{4-1/\eta}}.
\label{eq:dimergap}
\end{align}
In addition, the difference between the ground-state energy 
${\cal E}_{\rm free}$ of 
the free-boson theory~(\ref{eq:XXZ_eff_H}) with $\lambda=0$ 
per site and that of the sine-Gordon theory~(\ref{eq:XXZ_dimer_eff}), 
${\cal E}_{\rm SG}$, has been predicted as~\cite{Zamolodchikov95,Lukyanov97}
\begin{align}
\frac{\Delta{\cal E}_{\rm GS}}{J}=\frac{{\cal E}_{\rm free}-{\cal E}_{\rm SG}}{J}
=\frac{1}{4}\frac{v}{Ja}
\left(\frac{Ja}{v}\frac{E_{S}}{J}\right)^2\tan
\left(\frac{\pi}{2}\frac{1}{4\eta-1}\right).
\label{eq:E_GS}
\end{align}
However, we should note that the above formula is invalid for the 
ferromagnetic side $\Delta_z\leq 0$ ($\eta\leq 1/2$) 
since it diverges at the XY point $\Delta_z=0$ ($\eta=1/2$).

A similar sine-Gordon model also emerges
in spin-$\frac{1}{2}$ XXZ chains in a staggered field, 
\begin{equation}
{\cal H}^{\rm stag}={\cal H}^{\rm XXZ}+\sum_{j}(-1)^{j}h_{\rm s}S_{j}^{z}.
\label{eq:XXZ_stag}
\end{equation}
The staggered field $h_{\rm s}$ induces a relevant perturbation 
$\cos(\sqrt{4\pi}\phi)$. 
Therefore, the resultant effective Hamiltonian is 
\begin{equation}
{\cal H}_{\rm eff}^{\rm stag}={\cal H}_{\rm eff}^{\rm XXZ}+
\int{\rm d}x \frac{h_{\rm s}}{a}a_{1}
\cos(\sqrt{4\pi}\phi).
\label{eq:XXZ_stag_eff}
\end{equation}
If we redefine the scalar field $\phi$ as $\phi+\sqrt{\pi}/4$, 
the form of Eq.~(\ref{eq:XXZ_stag_eff}) becomes equivalent to 
that of Eq.~(\ref{eq:XXZ_dimer_eff}). 
Thus, the soliton gap of the model~(\ref{eq:XXZ_stag_eff}) is 
equal to Eq.~(\ref{eq:dimergap}) with the replacement 
of $\delta_{xy}d_{xy}+\delta_{z}d_{z}\to h_{\rm s}a_{1}/J$. 
Namely the soliton gap of the model~(\ref{eq:XXZ_stag_eff}) is given by
\begin{align}
\frac{E_S}{J}=&\frac{v}{Ja}\frac{2}{\sqrt{\pi}}
\frac{\Gamma\left(\frac{1}{8\eta-2}\right)}{\Gamma\left(\frac{2}{4-1/\eta}
\right)}
\nonumber\\
&\times\left[\frac{Ja}{v}\frac{\pi (h_{\rm s}a_{1})}{2J}
\frac{\Gamma\left(\frac{4-1/\eta}{4}\right)}
{\Gamma\left(\frac{1}{4\eta}\right)} \right]^{\frac{2}{4-1/\eta}}.
\label{eq:staggap}
\end{align}
This type of staggered-field induced gaps has been observed in some 
quasi 1D magnets with an alternating gyromagnetic tensor or 
Dzyaloshinskii-Moriya interaction such as 
Cu benzoate.~\cite{Oshikawa97,Affleck99,Essler98,Kuzmenko09,Dender97}


Masses of the soliton, antisoliton and breathers are related to 
the excitation gaps of 
the original lattice systems, Eqs.~(\ref{eq:dimer_XXZ}) and 
(\ref{eq:XXZ_stag}). The soliton and antisoliton 
correspond to the lowest excitations which change the $z$ component of 
total spin $S^z_{\rm tot}=\sum_jS_j^z$ by $\pm 1$. 
On the other hand, the lightest breather is regarded as the lowest 
excitation with $\Delta S^z_{\rm tot}=0$. 
At the $SU(2)$-symmetric AF point $\Delta_z=1$, there are 
three breathers. 
The soliton, antisoliton and lightest breather are degenerate and 
form the spin-1 triplet excitations (so-called magnons). 
The second lightest breather is interpreted as the singlet excitation 
with $\Delta S_{\rm tot}=0$. 
In the ferromagnetic regime $\Delta_z<0$, where any breather disappears, 
the lowest soliton-antisoliton scattering state 
would correspond to the excitation gap in the sector of 
$\Delta S_{\rm tot}^z=0$.

\subsection{${\bol J}$-${\bol J}_{\bol 2}$ antiferromagnetic spin chain}
\label{subsec:J-J2}
In the previous two subsections, we have completely 
neglected effects of irrelevant perturbations in 
the low-energy effective theory. However, 
as already noted, the $\lambda$ term becomes nearly 
marginal when the anisotropy $\Delta_z$ approaches unity. 
In this case, the $\lambda$ term is expected to affect 
several physical quantities. 
Actually, such effects have been studied in both 
the models~(\ref{eq:dimer_XXZ}) [Ref.~\onlinecite{Orignac04}] 
and (\ref{eq:XXZ_stag}) [Refs.~\onlinecite{Oshikawa97,Affleck99}].

It is known~\cite{Haldane82} that a small AF NNN coupling $J_2$ 
decreases the value of $\lambda$ in the $SU(2)$-symmetric AF 
Heisenberg chain. Okamoto and Nomura~\cite{Okamoto92} have shown that 
the marginal interaction vanishes, i.e., $\lambda\to 0$ 
in the following model: 
\begin{equation}
{\cal H}^{\rm nnn}=\sum_{j}(J\bol{S}_{j}\cdot\bol{S}_{j+1}
+J_{2}\bol{S}_{j}\cdot\bol{S}_{j+2}),
\label{eq:J-J2chain}
\end{equation}
with $J_2=0.2411J$. On the $J_2/J$ axis, this model is 
located at the Kosterlitz-Thouless transition point between 
the TLL and a spontaneously dimerized phase. 
From this fact, if we replace ${\cal H}^{\rm XXX}$ with 
${\cal H}^{\rm nnn}$ in the $SU(2)$-symmetric models~(\ref{eq:dimer_XXZ}) 
and (\ref{eq:XXZ_stag}), namely, if we consider the following models:
\begin{subequations}
\label{eq:modified}
\begin{align}
\tilde{\cal H}^{{\rm XXX}\mathchar`-\delta}
=&{\cal H}^{\rm nnn}+\sum_{j}(-1)^{j}\delta J\bol{S}_{j}\cdot\bol{S}_{j+1},
\label{eq:modified_dimer}\\
\tilde{\cal H}^{\rm stag}=&{\cal H}^{\rm nnn}+\sum_{j}(-1)^{j}h_{\rm s}S_{j}^{z},
\label{eq:modified_stag}
\end{align}
\end{subequations}
then their effective theories are much 
closer to a pure sine-Gordon model. 
In other words, the predictions from the sine-Gordon model, 
such as Eqs.~(\ref{eq:dimergap}) and (\ref{eq:staggap}), 
become more reliable.

\section{Coefficients of Dimer and Spin Operators}
\label{sec:delta}
From the discussions in Sec.~\ref{sec:dimer}, one can readily find a way 
of extracting the values of $d_{xy,z}$ and $a_1$ in Eqs.~(\ref{eq:dimer}) 
and (\ref{eq:spin_boson}) as follows. 
We first calculate some low-energy levels in 
$S_{\rm tot}^z=\pm1$ and $S_{\rm tot}^z=0$ sectors of the models 
(\ref{eq:dimer_XXZ}), (\ref{eq:XXZ_stag}) and (\ref{eq:modified}) 
by means of numerical diagonalization method. 
Since all the Hamiltonians~(\ref{eq:dimer_XXZ}), (\ref{eq:XXZ_stag}) and 
(\ref{eq:modified}) commute with $S^{z}_{\rm tot}=\sum_{j}S_{j}^{z}$, 
the numerical diagonalization can be performed in the Hilbert subspace 
with each fixed $S^{z}_{\rm tot}$. 
In order to extrapolate gaps to the thermodynamic limit 
with reasonable accuracy, we use appropriate finite-size scaling 
methods~\cite{Cardy84,Cardy86,Cardy86b,Shanks55} 
for spin chains under periodic boundary condition (total number of 
sites $L=8$, 10, $\cdots$, 28, 30). 
Secondly, the coefficients $d_{xy,z}$ and $a_1$ of 
the spin-$\frac{1}{2}$ XXZ chain and the $J$-$J_2$ chain are determined 
via the comparison between the sine-Gordon gap formula~(\ref{eq:dimergap}) 
and numerically evaluated spin gaps 
for various values of $\delta_{xy,z}$ and $h_{\rm s}$. 
In this procedure, (as already mentioned) the energy difference 
between the lowest (i.e., ground-state) and the second lowest levels 
of the $S_{\rm tot}^z=0$ sector (gap with $\Delta S_{\rm tot}^z=0$) and 
that between the ground-state level and the lowest level of 
the $S_{\rm tot}^z=\pm1$ sector (gap with $\Delta S_{\rm tot}^z=\pm1$) 
are respectively interpreted as the breather (or 
soliton-antisoliton scattering state) and soliton masses 
in the sine-Gordon scheme. 

\subsection{TLL phase and Numerical diagonalization} 
\label{subsec:TLLandED}
In this subsection, we focus on the TLL phase of uniform 
spin-$\frac{1}{2}$ XXZ chains~(\ref{eq:XXZ}) and 
test the reliability of our numerical diagonalization. 
The low-energy properties are described by Eq.~(\ref{eq:XXZ_eff_H}), 
which is a free boson theory (i.e., CFT with central charge $c=1$) 
with some irrelevant perturbations. 
Generally, the finite-size scaling formula for 
the excitation spectrum in any CFT has been proved~\cite{Cardy84,Cardy86} 
to be
\begin{align}
\Delta E_{\cal O}\equiv E_{\cal O}-E_{0}=\frac{2\pi v}{La} [{\cal O}]+\cdots.
\label{eq:finite_size}
\end{align}
Here $E_0$ and $E_{\cal O}$ are respectively the ground-state energy and 
the energy of an excited state generating from a primary field ${\cal O}$ 
in the given CFT. Remaining quantities $[{\cal O}]$, $v$, and $La$ are the 
scaling dimension of the operator ${\cal O}$, the excitation velocity 
and the system length, respectively. 
In the case of the spin chain~(\ref{eq:XXZ}), 
the bosonization formula~(\ref{eq:spin_boson}) indicates that 
$E_{e^{\pm i\sqrt{\pi}\theta}}$ and $E_{e^{\pm i\sqrt{4\pi}\phi}}$ 
correspond to the excitation energies in the $S_{\rm tot}^z=\pm 1$ 
and $S_{\rm tot}^z=0$ sectors, respectively. The irrelevant perturbations 
can also contribute to the finite-size 
correction to excitation energies. From the $U(1)$ and translational 
symmetries of the XXZ chain~(\ref{eq:XXZ}), one can show that the finite-size 
gap $\Delta E_{\Delta S_{\rm tot}^z=\pm 1}$ has no significant 
modification from the perturbations, while the correction to 
$\Delta E_{\Delta S_{\rm tot}^z= 0}$ is proportional 
to $L^{1-[{\rm e}^{{\rm i}2\sqrt{4\pi}\phi}]}$. 
Therefore, the following finite-size scaling formulas are predicted:
\begin{subequations}
\label{eq:gap_correction}
\begin{align}
\Delta E_{\Delta S_{\rm tot}^z=\pm 1}\approx
\frac{2\pi v}{La}\frac{1}{4K}+\cdots,\\ 
\Delta E_{\Delta S_{\rm tot}^z=0}\approx
\frac{2\pi v}{La}K+c_0 L^{1-4K}+\cdots,
\end{align}
\end{subequations}
with $c_0$ being a non-universal constant. 
Here we have used $[{\rm e}^{{\rm i}n\sqrt{\pi}\theta}]=n^2/(4K)$ 
and $[{\rm e}^{{\rm i}n\sqrt{4\pi}\phi}]=n^2K$. 
At the $SU(2)$-symmetric AF point $\Delta_z=1$, 
$\Delta E_{\Delta S_{\rm tot}^z=\pm 1}=\Delta E_{\Delta S_{\rm tot}^z=0}
\equiv \Delta E_{\rm su2}$ holds and the marginal $\lambda$ term modifies the 
scaling form of the spin gap. 
The marginal term is known to yield a logarithmic correction 
as follows:~\cite{Cardy86b}  
\begin{align}
\label{eq:gap_log}
\Delta E_{\rm su2}\approx\frac{2\pi v}{La}\left(\frac{1}{2}+
\frac{c_1}{\ln L}+\frac{c_2}{(\ln L)^2}+\cdots\right).
\end{align}
Here $c_{1,2}$ are non-universal constants.

\begin{figure}
\begin{center}
\includegraphics[width=0.35\textwidth]{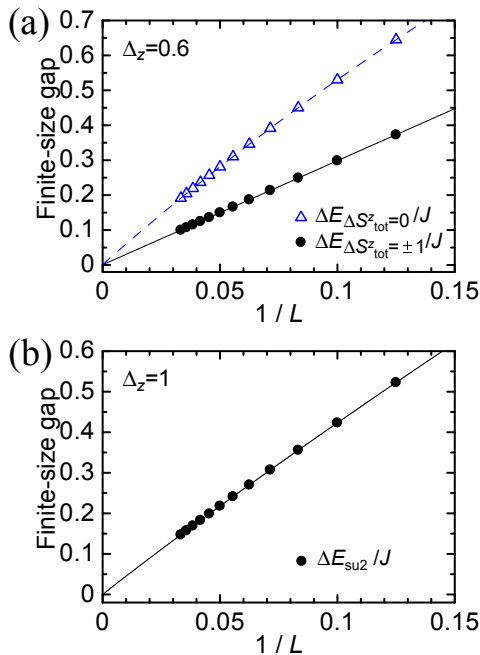}
\end{center}
\caption{(Color online) (a) Numerically evaluated gaps with 
$\Delta S^{z}_{\rm tot}=\pm1$ (circles) 
and $\Delta S^{z}_{\rm tot}=0$ (triangles) 
for XXZ chains with $\Delta_z=0.6$ and finite length $L$.
The solid curve $8.019\times10^{-4}+2.977/L$ (dashed curve 
$1.312\times10^{-3}+5.982/L-4.764/L^{1.8376}$) is 
determined by fitting the circles (triangles). (b) Gaps of finite-size 
Heisenberg chains with $\Delta_z=1$. The solid curve is 
$\Delta E_{\rm su2}/J=2.173\times 10^{-4}
+4.965/L-2.203/(L \ln L)+1.200/(L (\ln L)^2)$.}
\label{fig:Del0_6TLL}
\end{figure}

As an example, numerically evaluated gaps with 
$\Delta S^{z}_{\rm tot}=\pm1$ and $\Delta S^{z}_{\rm tot}=0$ 
in the case of $\Delta_z=0.6$ are 
respectively represented as circles and triangles in 
Fig.~\ref{fig:Del0_6TLL}(a). 
Circles are nicely fitted by the solid curve 
$\Delta E_{\Delta S_{\rm tot}^z=\pm 1}/J=8.019\times10^{-4}+2.977/L$. 
This result is consistent with the fact that
an easy-plane anisotropic XXZ model is gapless in the thermodynamic limit 
and that the exact coefficient of the $1/L$ term is $2\pi v/(4JK)=3$ at 
$\Delta_z=0.6$. Similarly, triangles can be fitted by  
$\Delta E_{\Delta S_{\rm tot}^z=0}/J=
1.312\times10^{-3}+5.982/L-4.764/L^{1.8376}$ where $1.8376=1-4K$. 
The factor 5.982 of the $1/L$ term is very close 
to $2\pi v K/(Ja)=6.040$. The spin gap at $SU(2)$-symmetric point 
is also represented in Fig.~\ref{fig:Del0_6TLL}(b). 
Following the formula~(\ref{eq:gap_log}), we can correctly determine 
the fitting curve $\Delta E_{\rm su2}/J=2.173\times 10^{-4}
+4.965/L-2.203/(L \ln L)+1.200/(L (\ln L)^2)$, 
in which the factor of the second term is nearly equal to 
$\pi v/(Ja)=4.935$. 
These results support the reliability of our numerical diagonalization.
We note that a more precise finite-size scaling analysis for 
AF Heisenberg model has been performed in Ref.~\onlinecite{Nomura93}.

\subsection{Dimer coefficients of XY model}
\label{subsec:XY_dimer}
Next, let us move onto the evaluation of excitation gaps in 
dimerized XXZ chains. In this case, since the system is not critical, 
the above finite-size scaling based on CFT cannot be applied.
Instead, we utilize Aitken-Shanks method~\cite{Shanks55} to extrapolate 
our numerical data to the values in the thermodynamic limit.

In this subsection, we consider a special dimerized XY chain 
with $\Delta_{z}=\delta_{z}=0$. It is mapped to a 
solvable free fermion system through Jordan-Wigner 
transformation. Therefore, our numerically determined coefficients in 
Eq.~(\ref{eq:dimer}) can be compared with the exact value. 
The lowest energy gap with $\Delta S_{\rm tot}^z=\pm 1$, 
which corresponds to the soliton mass $E_S$, is exactly evaluated as 
\begin{equation}
\Delta E_{\Delta S_{\rm tot}^z=\pm 1}/J=\delta_{xy}. \label{eq:XY_soliton}
\end{equation}
Comparing Eq.~(\ref{eq:XY_soliton}) with Eq.~(\ref{eq:dimergap}), 
we obtain the exact coefficient 
\begin{equation}
d_{xy}=1/\pi=0.3183
\end{equation}
at the XY case $\Delta_{z}=\delta_{z}=0$. The exact solution also tells us 
that the excitation gap with $\Delta S_{\rm tot}^z=0$ is 
\begin{align}
\Delta E_{\Delta S_{\rm tot}^z=0}/J=2\delta_{xy}. \label{eq:XY_breather}
\end{align}
This is consistent with the sine-Gordon prediction that any breather 
disappears and the relation $E_{B_1}=2E_S$ holds just 
at the XY point $\Delta_z=0$. 

\begin{figure}
\begin{center}
\includegraphics[width=0.35\textwidth]{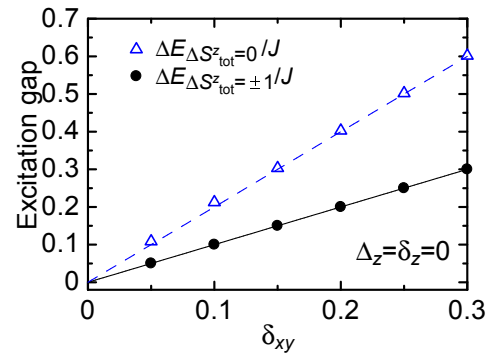}
\end{center}
\caption{(Color online) Numerically evaluated gaps 
$\Delta E_{\Delta S_{\rm tot}^z=\pm 1}$ (circles) and 
$\Delta E_{\Delta S_{\rm tot}^z=0}$ (triangles) in dimerized XY models with 
$\Delta_{z}=\delta_{z}=0$. Solid and dashed lines are 
the exact results determined via Jordan-Wigner transformation. 
These lines respectively correspond to the soliton and breather masses 
in the framework of sine-Gordon theory.}
\label{fig:gap_XY}
\end{figure}

Figure~\ref{fig:gap_XY} shows the comparison between the energy gap 
calculated by numerical diagonalization with Aitken-Shanks process 
and Eq.~(\ref{eq:XY_soliton}) [or Eq.~(\ref{eq:XY_breather})].
Except for $\Delta E_{\Delta S_{\rm tot}^{z}=0}$ 
in the weak dimerized regime $\delta_{xy}\alt 0.1$, 
numerically calculated gaps coincide well with the exact value. 
We have found that when $\delta_{xy,z}$ becomes smaller, 
the precision of Aitken-Shanks method is decreased due to a large 
size dependence of gaps.

\subsection{Dimer coefficients of XXZ model}
\label{subsec:XXZ_dimer}

\begin{table*}
\caption{\label{tb:dimer_coeff} Dimer coefficients ($d_{xy}$ and $d_z$), 
TLL parameter $K$, compactification radius $R$, spinon velocity $v$ 
of spin-$\frac{1}{2}$ XXZ chain. Dimerization-induced gaps are 
also listed in the final column. The final line is the result for 
the $J$-$J_2$ chain~(\ref{eq:J-J2chain}). The same data of $d_{xy,z}$ are also 
shown in Fig.~\ref{fig:dimer_coeff}.}
\begin{ruledtabular}
\begin{tabular}{lllllll}
\multicolumn{1}{c}{$\Delta_z$} 
& \multicolumn{1}{c}{$d_{xy}$} & 
\multicolumn{1}{c}{$d_z$} & \multicolumn{1}{c}{$K$} & \multicolumn{1}{c}{$R$} & 
\multicolumn{1}{c}{$v/(Ja)$} & \multicolumn{1}{c}{soliton gap $E_S/J$} \\
\hline
$1$       & 0.228 (0.204) & 0.110 (0.097) &  0.5    & 0.3989($=1/\sqrt{2\pi}$) & 1.571($=\pi/2$)  & $3.535(\delta_{xy}d_{xy}+\delta_{z}d_{z})^{0.6667}$\\
$0.9$     & 0.278 (0.261) & 0.141 (0.131) & 0.5838 & 0.3692 & 1.518  & $3.268(\delta_{xy}d_{xy}+\delta_{z}d_{z})^{0.7061}$\\
$0.8$     & 0.297 (0.284) & 0.154 (0.146) & 0.6288 & 0.3557 & 1.465  & $3.147(\delta_{xy}d_{xy}+\delta_{z}d_{z})^{0.7293}$\\
$0.7$     & 0.309 (0.299) & 0.165 (0.159) & 0.6695 & 0.3448 & 1.410  & $3.057(\delta_{xy}d_{xy}+\delta_{z}d_{z})^{0.7516}$\\
$0.6$     & 0.318 (0.310) & 0.174 (0.169) & 0.7094 & 0.3349 & 1.355  & $2.986(\delta_{xy}d_{xy}+\delta_{z}d_{z})^{0.7748}$\\
$0.5$     & 0.324 (0.318) & 0.182 (0.177) & 0.75   & 0.3257 & 1.299  & $2.934(\delta_{xy}d_{xy}+\delta_{z}d_{z})^{0.8}$\\
$0.4$     & 0.327 (0.323) & 0.188 (0.185) & 0.7924 & 0.3169 & 1.242  & $2.902(\delta_{xy}d_{xy}+\delta_{z}d_{z})^{0.8281}$\\
$0.3$     & 0.328 (0.325) & 0.193 (0.191) & 0.8375 & 0.3082 & 1.184  & $2.893(\delta_{xy}d_{xy}+\delta_{z}d_{z})^{0.8602}$\\
$0.2$     & 0.328 (0.325) & 0.197 (0.196) & 0.8864 & 0.2996 & 1.124  & $2.918(\delta_{xy}d_{xy}+\delta_{z}d_{z})^{0.8980}$\\
$0.1$     & 0.324 (0.323) & 0.200 (0.200) & 0.9401 & 0.2910 & 1.063  & $2.991(\delta_{xy}d_{xy}+\delta_{z}d_{z})^{0.9434}$\\
$0$       & 0.318 (0.318) & 0.202 (0.203) & 1      & 0.2821($=1/\sqrt{4\pi}$) & 1      & $3.141(\delta_{xy}d_{xy}+\delta_{z}d_{z})$\\
$-0.1$    & 0.309 (0.311) & 0.202 (0.204) & 1.068  & 0.2730 & 0.9353 & $3.431(\delta_{xy}d_{xy}+\delta_{z}d_{z})^{1.073}$\\
$-0.2$    & 0.297 (0.302) & 0.200 (0.204) & 1.147  & 0.2634 & 0.8685 & $4.008(\delta_{xy}d_{xy}+\delta_{z}d_{z})^{1.172}$\\
$-0.3$    & 0.278 (0.289) & 0.194 (0.203) & 1.241  & 0.2533 & 0.7990 & $5.308(\delta_{xy}d_{xy}+\delta_{z}d_{z})^{1.317}$\\
$-0.4$    & 0.252 (0.273) & 0.184 (0.199) & 1.355  & 0.2423 & 0.7263 & $9.214(\delta_{xy}d_{xy}+\delta_{z}d_{z})^{1.550}$\\
$-0.5$    & 0.213 (0.248) & 0.163 (0.191) & 1.5    & 0.2303 & 0.6495 & $33.25(\delta_{xy}d_{xy}+\delta_{z}d_{z})^{2}$\\
$J$-$J_{2}$ model & 0.364 (0.361) & 0.188 (0.182) & 0.5    & 0.3989 & 1.174 & $3.208(\delta_{xy}d_{xy}+\delta_{z}d_{z})^{0.6667}$
\end{tabular}
\end{ruledtabular}
\end{table*}

\begin{figure}
\begin{center}
\includegraphics[width=0.4\textwidth]{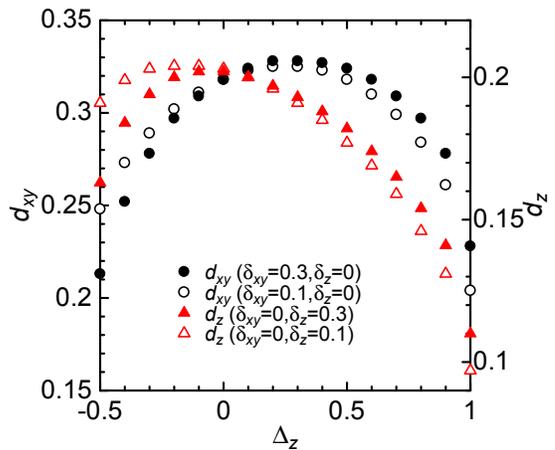}
\end{center}
\caption{(Color online) XXZ-anisotropy ($\Delta_z$) dependence of 
dimer coefficients $d_{xy}$ and $d_z$. Filled [Open] circles represent 
$d_{xy}$ determined from dimerization gap at 
$(\delta_{xy},\delta_z)=(0.3,0)$ [$=(0.1,0)$]. Similarly, 
filled [open] triangles show $d_z$ determined from 
dimerization gap at $(\delta_{xy},\delta_z)=(0,0.3)$ [$=(0,0.1)$].} 
\label{fig:dimer_coeff}
\end{figure}

In the easy-plane region $-1<\Delta_{z}<1$, 
any generic analytical way of determining the coefficients 
in Eq.~(\ref{eq:dimer}) has never been known except for 
the above special point $\Delta_z=\delta_z=0$. 
To obtain $d_{xy}$ (respectively $d_{z}$), we numerically 
calculate excitation gaps 
at the points $\delta_{xy}$ $(\delta_{z})=0.05$, 
0.1, $\cdots$, 0.3 with fixing $\delta_{z}(\delta_{xy})=0$. 
Although both $\Delta E_{\Delta S_{\rm tot}^{z}=0}$ and 
$\Delta E_{\Delta S_{\rm tot}^{z}=\pm 1}$ are applicable to determine 
$d_{xy,z}$ in principle, 
we use only the latter gap since it more smoothly converges 
to its thermodynamic-limit value via Aitken-Shanks process, 
compared to the former. 
In fact, Eq.~(\ref{eq:gap_correction}) suggests that 
$\Delta E_{\Delta S_{\rm tot}^{z}=0}$ is subject to effects of 
irrelevant perturbations and therefore contains complicated 
finite-size corrections. 
Coefficients $d_{xy}$ ($d_{z}$) can be determined for each 
$\delta_{xy}$ ($\delta_{z}$) from Eq.~(\ref{eq:dimergap}). 
Since the field theory result~(\ref{eq:dimergap}) is generally 
more reliable as the perturbation $\delta_{xy,z}$ is smaller, 
we should compare Eq.~(\ref{eq:dimergap}) with excitation gaps 
determined at sufficiently small values of $\delta_{xy,z}$. 
However, the extrapolation to thermodynamic limit by Aitken-Shanks method
is less precise in such a small dimerization region 
mainly due to large finite-size effects.~\cite{Papenbrock03,Orignac04}
Therefore, we adopt coefficients $d_{xy,z}$ extracted from the gaps at 
relatively large dimerization $\delta_{xy(z)}=0.1$ and $0.3$, and 
they are listed in Table~\ref{tb:dimer_coeff}: the values outside 
[inside] parentheses are the data for 
$\delta_{xy(z)}=0.3$ [0.1]. The anisotropy dependence of 
the same data $d_{xy,z}$ is depicted in Fig.~\ref{fig:dimer_coeff}. 
The data in Table~\ref{tb:dimer_coeff} and Fig.~\ref{fig:dimer_coeff} 
are the main result of this paper. The difference 
between $d_{xy(z)}$ outside and inside the parentheses 
in Table~\ref{tb:dimer_coeff} could be interpreted as 
the "strength" of irrelevant perturbations neglected in the effective 
sine-Gordon theory or the "error" of our numerical strategy. 
The neglected operators must 
bring a renormalization of coefficients $d_{xy,z}$, and the "error" 
would become larger as the system approaches the Heisenberg point 
since (as already mentioned) the $\lambda$ term becomes marginal 
at the point.

We here discuss the validity of the numerically determined 
$d_{xy,z}$ in Table~\ref{tb:dimer_coeff} and Fig.~\ref{fig:dimer_coeff}. 
Table~\ref{tb:dimer_coeff} shows that in the wide range 
$-0.3\alt \Delta_z\alt0.9$, the difference (error) between $d_{xy,z}$ 
outside and inside the parentheses is less than 8 $\%$. As expected, 
one finds that the error gradually increases when the anisotropy 
$\Delta_z$ approaches unity. Similarly, the error is large in 
the deeply ferromagnetic regime $\Delta_z\alt-0.3$. 
This is naturally understood from the fact that as $\Delta_z$ is 
negatively increased, the dimerization term $\sin(\sqrt{4\pi}\phi)$ 
becomes less relevant and effects of other irrelevant terms is 
relatively strong. Indeed, for $\Delta_z<-0.7071$ ($K>2$), 
the dimerization does not yield any spin gap and our method of 
determining $d_{xy,z}$ cannot be used. 
Furthermore, it is worth noting that the spin gap is 
convex downward as a function of dimerization $\delta_{xy,z}$ 
in the ferromagnetic side $\Delta_z<0$, and 
the accuracy of the fitting therefore depreciates. 

\begin{figure}
\begin{center}
\includegraphics[width=0.35\textwidth]{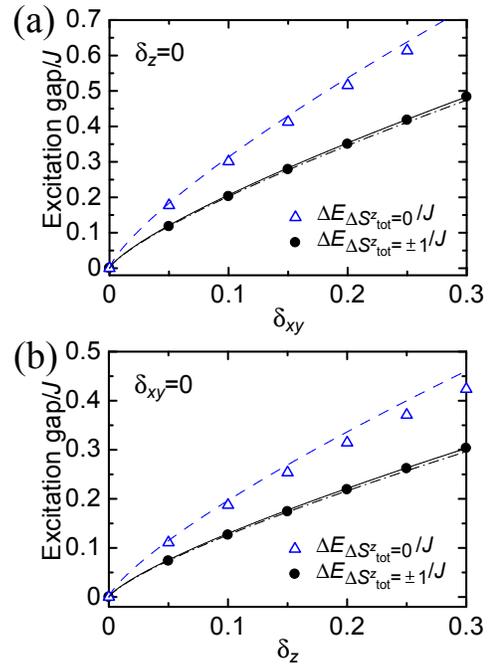}
\end{center}
\caption{(Color online) Excitation gaps 
$\Delta E_{\Delta S_{\rm tot}^{z}=\pm 1}$ (circles) and 
$\Delta E_{\Delta S_{\rm tot}^{z}=0}$ (triangles) 
of the dimerized XXZ model~(\ref{eq:dimer_XXZ}) with $\Delta_z=0.6$. 
Solid and dashed-dotted curves are fitted by the gap 
formula~(\ref{eq:dimergap}) with coefficients outside and 
inside parentheses in Table~\ref{tb:dimer_coeff}, respectively. 
The dashed curve represents the lightest breather mass which is fixed 
by combining the solid soliton curve and Eq.~(\ref{eq:mass_relation}).} 
\label{fig:gap_Del0_6}
\end{figure}

\begin{figure}
\begin{center}
\includegraphics[width=0.35\textwidth]{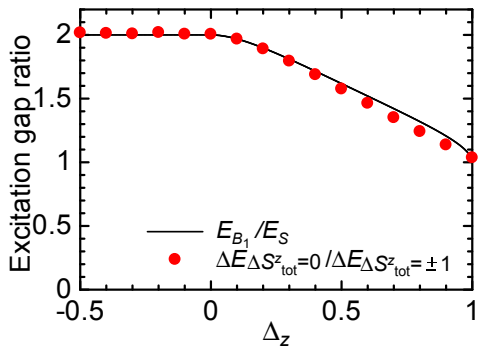}
\end{center}
\caption{(Color online) Ratio between two numerically evaluated gaps 
$\Delta E_{\Delta S_{\rm tot}^{z}=0}/\Delta E_{\Delta S_{\rm tot}^{z}=\pm1}$
(circles) in the dimerized chain~(\ref{eq:dimer_XXZ}) 
with $\delta_{xy}=0.3$ and $\delta_z=0$. Solid curve is the soliton-breather 
mass ratio $E_{B_1}/E_S$ in the effective sine-Gordon 
theory~(\ref{eq:XXZ_dimer_eff}). Note that in the ferromagnetic side 
$\Delta_z<0$, there is no breather and $E_{B_1}$ is replaced with 
the mass gap of soliton-antisoliton scattering states $2E_S$, 
namely, $E_{B_1}/E_S\to 2E_S/E_S=2$.}
\label{fig:compare_ED_SG}
\end{figure}

\begin{figure}
\begin{center}
\includegraphics[width=0.35\textwidth]{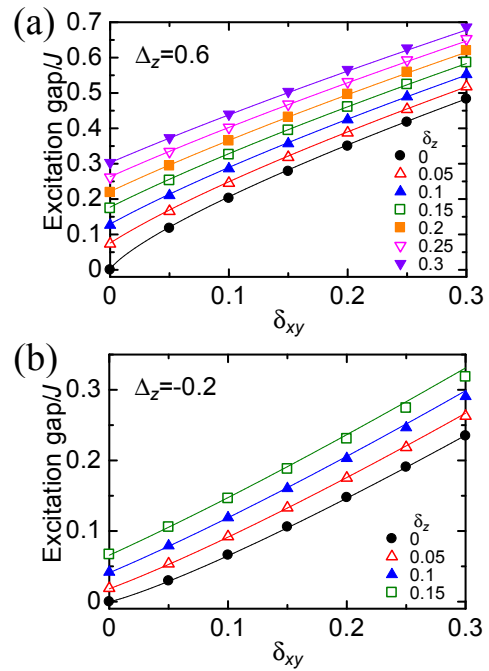}
\end{center}
\caption{(Color online) Numerically evaluated gaps 
$\Delta E_{\Delta S_{\rm tot}^{z}=\pm1}$ of 
dimerized XXZ chains with several values of 
both parameters $\delta_{xy}$ and $\delta_z$ at $\Delta_z=0.6$ and $-0.2$. 
Solid curves are Eq.~(\ref{eq:dimergap}) with $d_{xy}$ and $d_z$ 
in Table~\ref{tb:dimer_coeff}. In the ferromagnetic case $\Delta_z=-0.2$, 
the analytical curve successfully fits the numerical data for a wide 
weakly-dimerized regime $\delta_{xy,z}\ll 1$, 
while the deviation occurs for the strongly-dimerized one.}
\label{fig:gap_Del0_6all}
\end{figure}

In addition to coefficients $d_{xy,z}$, 
let us examine dimerization gaps and 
the quality of fitting by Eq.~(\ref{eq:dimergap}). 
Excitation gaps for $\Delta_{z}=0.6$ are shown 
in Fig.~\ref{fig:gap_Del0_6} as an example.
Remarkably, both soliton-gap curves~(\ref{eq:dimergap}) with the values 
$d_{xy,z}$ outside and inside the parentheses in Table~\ref{tb:dimer_coeff} 
fit the numerical data $\Delta E_{\Delta S_{\rm tot}^{z}=\pm 1}$ 
in the broad region $0\leq\delta_{xy(z)}\leq 0.3$ with reasonable accuracy. 
The former solid curve is slightly better that the latter.  
The breather gaps $\Delta E_{\Delta S_{\rm tot}^{z}=0}$ and 
corresponding fitting curves 
are also shown in Fig.~\ref{fig:gap_Del0_6}. 
This breather curve is determined by combining the solid 
curve~(\ref{eq:dimergap}) and the soliton-breather 
relation~(\ref{eq:mass_relation}). 
It slightly deviates from numerical data, especially, 
in a relatively large dimerization regime $0.15\alt \delta_{xy(z)}$. 
As mentioned above, this deviation would be attributed to 
irrelevant perturbations. 
The breather-soliton mass ratio $E_{B_{1}}/E_{S}$ 
[see Eq.~(\ref{eq:mass_relation})] in the sine-Gordon 
model~(\ref{eq:XXZ_dimer_eff}) and the numerically evaluated 
$\Delta E_{\Delta S_{\rm tot}^{z}=0}/\Delta E_{\Delta S_{\rm tot}^{z}=\pm1}$ 
are shown in Fig.~\ref{fig:compare_ED_SG}. 
These two values are in good agreement with each other 
in the wide parameter region 
$-0.5<\Delta_{z}< 1$, although their difference becomes slightly 
larger in the region $0.5\alt\Delta_{z}\alt 1$, which includes 
the point $\Delta_z=0.6$ in Fig.~\ref{fig:gap_Del0_6}. 
Gaps $\Delta E_{\Delta S_{\rm tot}^{z}=\pm 1}$ for dimerized XXZ 
chains with several values of both $\delta_{xy}$ and $\delta_z$ are plotted 
in Fig.~\ref{fig:gap_Del0_6all}. It shows that the numerical data are 
quantitatively fitted by the {\it single} gap formula~(\ref{eq:dimergap}). 
All of the results in Figs.~\ref{fig:gap_Del0_6}-\ref{fig:gap_Del0_6all} 
indicates that a simple sine-Gordon model~(\ref{eq:XXZ_dimer_eff}) can 
describe the low-energy physics of the dimerized spin 
chain~(\ref{eq:dimer_XXZ}) 
with reasonable accuracy in the wide easy-plane regime. 
This also supports the validity of our 
numerical approach for fixing the coefficients $d_{xy,z}$.

\subsection{Dimer coefficients of SU(2)-symmetric models}
\label{subsec:SU2_dimer}
\begin{figure}
\begin{center}
\includegraphics[width=0.35\textwidth]{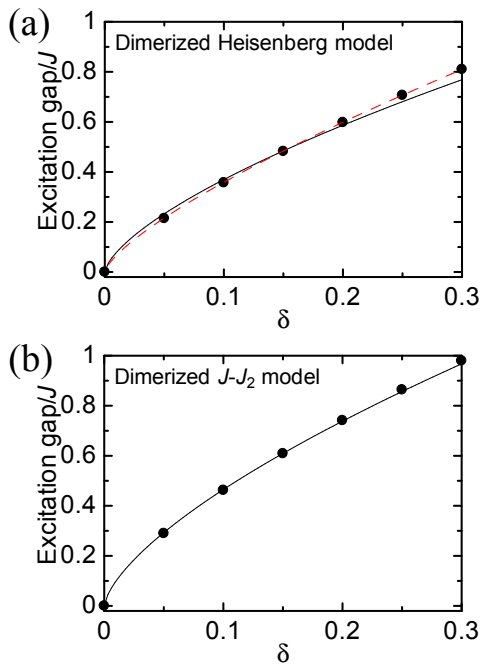}
\end{center}
\caption{(Color online) Spin gaps (circles) of (a) the Heisenberg model 
with dimerization $\delta_{xy}=\delta_{z}=\delta$ and 
(b) the dimerized $J$-$J_{2}$ model~(\ref{eq:modified_dimer}).
Both solid curves in panels (a) and (b) are determined from 
the gap formula~(\ref{eq:dimergap}). The dashed curve in panel (a) 
represents Eq.~(\ref{eq:gap_SU2_dimer_corr}).} 
\label{fig:gap_Heisen_dimer}
\end{figure}

At the $SU(2)$-symmetric AF point, 
the $\lambda$ term in the effective Hamiltonian~(\ref{eq:XXZ_eff_H}) 
becomes marginal and induces logarithmic corrections to several 
physical quantities. Such a logarithmic fashion often makes the accuracy 
of numerical methods decrease. 
Instead of numerical approaches, using the asymptotic form 
of the spin correlation function~\cite{Affleck98} 
and OPE technique,~\cite{Gogolin,Tsvelik} 
Orignac~\cite{Orignac04} has predicted 
\begin{equation}
\label{eq:dimer_SU2}
d_{xy}=2d_z = \frac{2}{\pi^{2}}\left(\frac{\pi}{2}\right)^{1/4}=0.2269
\end{equation}
at the $SU(2)$-symmetric point. Substituting Eq.~(\ref{eq:dimer_SU2}) 
into Eq.~(\ref{eq:dimergap}), 
the spin gap in a $SU(2)$-symmetric AF chain with dimerization 
$\delta_{xy}=\delta_z\equiv\delta$ (${\cal H}^{{\rm XXX}\mathchar`-\delta}$) 
is determined as
\begin{align}
\label{eq:gap_SU2_dimer}
\Delta E_{\rm su2}/J = 1.723 \delta^{2/3}.
\end{align}
The marginal term however produces a correction to this result. 
It has been shown in Ref.~\onlinecite{Orignac04} that the 
spin gap in the model ${\cal H}^{{\rm XXX}\mathchar`-\delta}$ 
is more nicely fitted with
\begin{equation}
\label{eq:gap_SU2_dimer_corr}
\Delta E_{\rm su2}/J = \frac{1.723\delta^{2/3}}
{\left(1+0.147\ln\Big|\frac{0.1616}{\delta}\Big|\right)^{1/2}},
\end{equation}
from the renormalization-group argument. 
As can be seen from Eq. (\ref{eq:gap_SU2_dimer_corr}), 
the logarithmic correction is not significantly large 
for the spin gap. We may therefore apply the way based on the 
sine-Gordon model in Sec.~\ref{subsec:XXZ_dimer} 
even for the present AF Heisenberg model. The resultant data are listed in 
the first line of Table~\ref{tb:dimer_coeff}. 
Evaluated coefficients $d_{xy}=0.228$ (0.204) and $d_{z}=0.110$ (0.097) 
are fairly close to the results of Eq.~(\ref{eq:dimer_SU2}). 
This suggests that the effect of the marginal operator on 
the spin gap is really small. 
We should also note that $d_{xy}= 2d_z$ is approximately 
realized, which is required from the $SU(2)$ symmetry. 
The numerically calculated spin gap $\Delta E_{\rm su2}$, 
Eq.~(\ref{eq:gap_SU2_dimer_corr}), and the curve of the gap 
formula~(\ref{eq:dimergap}) 
are shown in Fig.~\ref{fig:gap_Heisen_dimer}(a). It is found that 
even the curve without any logarithmic correction 
can fit the numerical data within semi-quantitative level. 
At least, parameters $d_{xy,z}$ at the $SU(2)$-symmetric point can 
be regarded as effective coupling constants when we naively approximate 
a dimerized Heisenberg chain as a simple sine-Gordon model.

As discussed in Sec.~\ref{subsec:J-J2}, logarithmic corrections vanish 
in the $J$-$J_{2}$ model~(\ref{eq:J-J2chain}) due to the absence of 
the marginal operator. As expected, Fig.~\ref{fig:gap_Heisen_dimer}(b) 
shows that the spin gap $\Delta E_{\rm su2}$ is accurately fitted by 
the sine-Gordon gap formula~(\ref{eq:dimergap}) in the wide range 
$0\leq\delta\leq0.3$. Therefore, the coefficients $d_{xy,z}$ of the $J$-$J_2$ 
model (the final line of Table~\ref{tb:dimer_coeff}) are highly reliable. 
Remarkably, the difference between the values 
outside and inside the parentheses is much smaller than that 
of the Heisenberg model (the first and last line of 
Table~\ref{tb:dimer_coeff}).  
Here, to determine $d_{xy,z}$ of the $J$-$J_{2}$ model, 
we have used its spinon velocity $v=1.174Ja$, 
which has been evaluated in Ref.~\onlinecite{Okamoto97}.

\subsection{Coefficients of spin operator}
\label{subsec:spin_op}
In this subsection, we discuss the spin-operator coefficient $a_{1}$ 
in Eq.~(\ref{eq:spin_boson}). 
Although $a_1$ for the easy-plane XXZ model has been evaluated 
analytically~\cite{Lukyanov97,Lukyanov99,Lukyanov03} and 
numerically,~\cite{Hikihara98,Hikihara04} those for 
the $SU(2)$-symmetric Heisenberg chain and the $J$-$J_2$ model 
have never been studied. The existent data also help us to check 
the validity of our method. 
From the bosonization formula~(\ref{eq:spin_boson}), the $z$-component 
spin correlation function has the following asymptotic form:  
\begin{align}
\langle S_{j}^{z}S_{j'}^{z}\rangle=
-\frac{1}{4\pi^{2}\eta|j-j'|^{2}}+
\frac{A_{1}^{z}(-1)^{j-j'}}{|j-j'|^{1/\eta}}+\cdots,
\end{align}
in the easy-plane TLL phase. 
The amplitude $A_1^z$ is related to $a_1$ as
\begin{align}
A_{1}^{z}=a_{1}^{2}/2.
\label{eq:amp_coeff}
\end{align}
Lukyanov and his collaborators \cite{Lukyanov97,Lukyanov99} have predicted 
\begin{align}
&A_{1}^{z}=\frac{2}{\pi^{2}}\left[\frac{\Gamma(\frac{\eta}{2-2\eta})}
{2\sqrt{\pi}\Gamma(\frac{1}{2-2\eta})}\right]^{1/\eta}\nonumber\\
&\times\exp\left[\int_{0}^{\infty}\frac{{\rm d}t}{t}
\left(\frac{\sinh[(2\eta-1)t]}{\sinh(\eta t)\cosh[(1-\eta)t]}-\frac{2\eta-1}{\eta}
{\rm e}^{-2t}\right)\right].
\label{eq:a1_anal}
\end{align}
The same amplitude has been calculated by using DMRG in 
Refs.~\onlinecite{Hikihara98,Hikihara04}. 

\begin{table*}
\caption{\label{tb:a1} Spin-operator coefficients $a_1$ 
of spin-$\frac{1}{2}$ XXZ chain and the $J$-$J_2$ chain. 
Values in column (A), (B), and (C) correspond to the analytical prediction 
from Refs.~\onlinecite{Lukyanov97,Lukyanov99,Lukyanov03}, the result by 
DMRG in Refs.~\onlinecite{Hikihara98,Hikihara04}, and ours, respectively.}
\begin{ruledtabular}
\begin{tabular}{lllllll}
\multicolumn{1}{c}{$\Delta_z$} & 
\multicolumn{1}{c}{$a_1$ (A)
} & 
\multicolumn{1}{c}{$a_1$ (B)
} & 
\multicolumn{1}{c}{$a_1$ (C) 
} 
& \multicolumn{1}{c}{$\eta$} &
\multicolumn{1}{c}{$v/(Ja)$} & \multicolumn{1}{c}{soliton gap $E_S/J$} \\
\hline
1   &        &        & 0.4724 (0.4325) & 1      & 1.571 & $3.535(a_{1}h_{\rm s}/J)^{0.6667}$\\
0.9 & 0.7049 & 0.64   & 0.5327 (0.4830) & 0.8564 & 1.518 & $3.268(a_{1}h_{\rm s}/J)^{0.7061}$\\
0.8 & 0.6069 & 0.587  & 0.5226 (0.4808) & 0.7952 & 1.465 & $3.147(a_{1}h_{\rm s}/J)^{0.7293}$\\
0.7 & 0.5464 & 0.54   & 0.5019 (0.4693) & 0.7468 & 1.410 & $3.057(a_{1}h_{\rm s}/J)^{0.7516}$\\
0.6 & 0.5008 & 0.499  & 0.4771 (0.4530) & 0.7048 & 1.355 & $2.986(a_{1}h_{\rm s}/J)^{0.7748}$\\
0.5 & 0.4629 & 0.4626 & 0.4505 (0.4338) & 0.6667 & 1.299 & $2.934(a_{1}h_{\rm s}/J)^{0.8}$   \\
0.4 & 0.4297 & 0.4297 & 0.4235 (0.4127) & 0.6310 & 1.242 & $2.902(a_{1}h_{\rm s}/J)^{0.8281}$\\
0.3 & 0.3994 & 0.3995 & 0.3966 (0.3903) & 0.5970 & 1.184 & $2.893(a_{1}h_{\rm s}/J)^{0.8602}$\\
0.2 & 0.3712 & 0.3713 & 0.3701 (0.3670) & 0.5641 & 1.124 & $2.918(a_{1}h_{\rm s}/J)^{0.8980}$\\
0.1 & 0.3443 & 0.3443 & 0.3440 (0.3430) & 0.5319 & 1.063 & $2.991(a_{1}h_{\rm s}/J)^{0.9434}$\\
0   & 0.3183 & 0.3183 & 0.3183 (0.3183) & 0.5    & 1     & $3.141(a_{1}h_{\rm s}/J)$         \\
$J$-$J_2$ model &  &  & 0.4693 (0.4668) & 1      & 1.174 & $3.208(a_{1}h_{\rm s}/J)^{0.6667}$\\
\end{tabular}
\end{ruledtabular}
\end{table*}

\begin{figure}
\begin{center}
\includegraphics[width=0.35\textwidth]{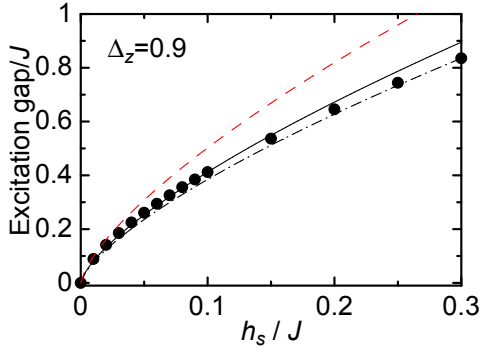}
\end{center}
\caption{(Color online) Spin gaps 
$\Delta E_{\Delta S_{\rm tot}^{z}=\pm1}$ (circles) of 
the XXZ models in a staggered field at $\Delta_z=0.9$. 
Solid and dashed-dotted curves are determined from the gap 
formula~(\ref{eq:staggap}) with $a_1$ outside and inside 
the parentheses in Table~\ref{tb:a1}, respectively. 
The dashed curve is given by the formula~(\ref{eq:staggap}) with 
$a_1$ in Refs.~\onlinecite{Lukyanov97,Lukyanov99,Lukyanov03} 
[i.e., $a_1$ determined from Eq.~(\ref{eq:a1_anal})]. } 
\label{fig:gap_Del0_9_stag}
\end{figure}

In order to determine $a_{1}$, we use XXZ models 
in a staggered field~(\ref{eq:XXZ_stag}). 
Following the similar way to Sec.~\ref{subsec:XXZ_dimer}, 
we can extract the coefficient $a_1$ by fitting numerically evaluated 
gaps of the model~(\ref{eq:XXZ_stag}) 
through the sine-Gordon gap formula~(\ref{eq:staggap}). 
We numerically estimate the gaps at $h_{\rm s}/J=0.01$, 0.02, $\cdots$, 0.09, 
0.1, 0.2, and 0.3 via Aitken-Shanks method. 
The results are listed in column (C) of Table~\ref{tb:a1}. 
Similarly to the case of dimerization, we adopt spin gaps at 
relatively large staggered fields $h_s/J=0.1$ and $0.3$ to determine 
the coefficients $a_1$. The value outside (inside) the parentheses 
in Table~\ref{tb:a1} corresponds to $a_1$ fixed at $h_{\rm s}/J=0.1$ (0.3). 
Note that the XY model in a staggered field is solvable through 
Jordan-Wigner transformation, and as a result 
the coefficient $a_1$ is exactly evaluated as
\begin{equation}
\label{eq:XY_stagg}
a_1 = 1/\pi=0.3183.
\end{equation}
The table clearly shows that the values 
at $h_{\rm s}/J=0.1$ are closer to those of the previous prediction in 
Refs.~\onlinecite{Lukyanov97,Lukyanov99,Lukyanov03,Hikihara98,Hikihara04}.
We emphasize that our results gradually deviate from the analytical 
prediction from Eq.~(\ref{eq:a1_anal}) as the system approaches 
the $SU(2)$-symmetric point. The same property also appears 
in the DMRG results in Refs.~\onlinecite{Hikihara98,Hikihara04}. 
Actually, $A_1^z$ in Eq.~(\ref{eq:a1_anal}) diverges when $\Delta_z\to1$. 
However, the bosonization formula~(\ref{eq:spin_boson}) for spin 
operators must be still used even around $\Delta_z=1$. 
Thus we should realize that the relation~(\ref{eq:amp_coeff}) is 
broken and $a_1$ remains to be finite at the $SU(2)$-symmetric point. 
Figure~\ref{fig:gap_Del0_9_stag} represents the numerically evaluated 
gaps $\Delta E_{\Delta S_{\rm tot}^{z}=\pm1}$, and three fitting curves 
fixed by $a_1$ (A) and $a_1$ (C) outside and inside the parentheses in 
Table~\ref{tb:a1}. Our coefficient $a_1$ successfully fits the numerical 
data semi-quantitatively in the wide regime $0.01\alt h_{\rm s}/J\alt 0.3$, 
while the curve of $a_1$ (A) is valid only in an extremely weak 
staggered-field regime $0<h_{\rm s}/J\alt0.01$. 
This implies that when $\Delta_{z}$ is near unity, 
the field theory description based on Eqs.~(\ref{eq:amp_coeff}) 
and (\ref{eq:a1_anal}) is valid only in a quite 
narrower region for the present staggered-field case 
compared to the case of dimerized spin chain. On the other hand, 
Fig.~\ref{fig:gap_Del0_9_stag} also suggests that if we use $a_1$ (C) in 
Table~\ref{tb:a1} as the effective coefficient of bosonized spin operator 
instead of $a_1$ (A) and (B), the XXZ chain in a 
staggered field (\ref{eq:XXZ_stag}) may be approximated by a simple 
sine-Gordon model in wide region $0.01\alt h_{\rm s}/J\alt 0.3$.

\begin{figure}
\begin{center}
\includegraphics[width=0.35\textwidth]{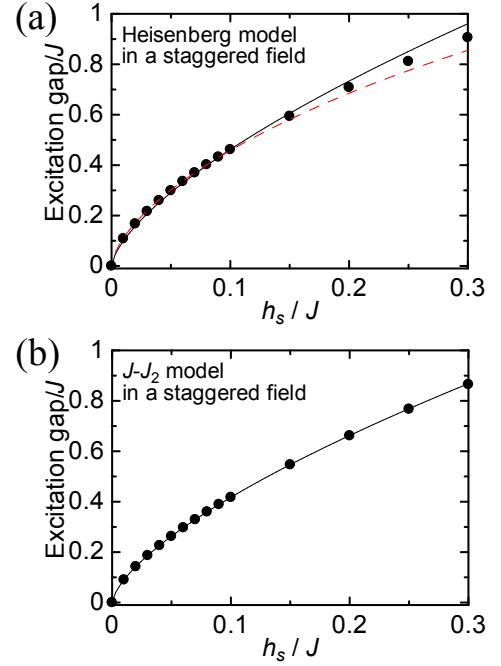}
\end{center}
\caption{(Color online) Spin gaps $\Delta E_{\Delta S_{\rm tot}^{z}=\pm1}$ 
(circles) of (a) the Heisenberg 
and (b) the $J$-$J_{2}$ models~(\ref{eq:modified_stag}) in a staggered field. 
Solid and dashed curves represent Eq.~(\ref{eq:staggap}) and 
Eq.~(\ref{eq:gap_SU2_stag_corr}), respectively. 
We have used $a_1$ outside the parentheses in column (C) of Table~\ref{tb:a1}. 
} 
\label{fig:gap_Heisen_stag}
\end{figure}

At the $SU(2)$-symmetric point $\Delta_{z}=1$, a logarithmic correction 
to staggered-field induced gaps is expected to appear due to the marginal 
perturbation. This makes it difficult to extract the value $a_1$ 
within the present sine-Gordon framework.
According to the prediction in Ref.~\onlinecite{Orignac04} 
based on the asymptotic form of spin correlation function,~\cite{Affleck98} 
$a_{1}$ is given by
\begin{equation}
\label{eq:spin_SU2}
a_1 = \frac{1}{\pi}\Big(\frac{\pi}{2}\Big)^{1/4}=0.3564
\end{equation}
at the $SU(2)$-symmetric point, where $a_1=b_0$ is imposed. 
The spin gap in AF Heisenberg chains in a staggered field 
(${\cal H}^{{\rm stag}}$ with $\Delta_z=1$) is thus determined as 
\begin{align}
\label{eq:gap_SU2_stag}
\Delta E_{\Delta S_{\rm tot}^{z}=\pm1}/J = 1.777 (h_{\rm s}/J)^{2/3}.
\end{align}
A more correct gap formula including the logarithmic correction 
has been developed in Refs.~\onlinecite{Oshikawa97,Affleck99} as follows:
\begin{equation}
\label{eq:gap_SU2_stag_corr}
\Delta E_{\Delta S_{\rm tot}^{z}=\pm1}/J = 
1.85 (h_{\rm s}/J)^{2/3}[\ln(J/h_{\rm s})]^{1/6}.
\end{equation}
In Fig.~\ref{fig:gap_Heisen_stag}(a), the numerically evaluated 
spin gaps, Eq.~(\ref{eq:gap_SU2_stag_corr}), and the fitting curve with 
$a_1$ outside the parentheses in column (C) are drawn. 
One finds that both curves agree well with the 
numerical data in the weak-field regime $0<h_{\rm s}/J\alt0.1$, 
while they start to deviate from the data in the stronger-field regime. 
This suggests that even at the $SU(2)$-symmetric point, a simple 
sine-Gordon description for the model~(\ref{eq:XXZ_stag}) is applicable 
in the relatively wide region $0<h_{\rm s}/J\alt0.1$, 
if the coefficient $a_1$ outside the parentheses in column (C) is adopted.

In the same way as the final paragraph in Sec.~\ref{subsec:SU2_dimer}, 
we can accurately determine the coefficient $a_1=b_0$ for the 
$J$-$J_{2}$ model 
since the marginal perturbation vanishes. 
The data are listed in the final line in Table~\ref{tb:a1}. 
One sees from Fig.~\ref{fig:gap_Heisen_stag}(b) that the spin 
gap $\Delta E_{\Delta S_{\rm tot}^{z}=\pm1}$ is fitted by the 
gap formula~(\ref{eq:staggap}) quite accurately. 
In addition, the difference between the values outside and inside 
the parentheses is significantly small.

\subsection{Coefficients determined from ground-state energy}
\label{subsec:gs}
Instead of the gap formula~(\ref{eq:dimergap}), 
the formula for ground-state energy~(\ref{eq:E_GS}) 
can also be utilized to determine dimer coefficients $d_{xy,z}$. 
Let us here define $\Delta E_{\rm GS}\equiv 
E_{\rm GS}-E_{\rm GS}(\delta_{xy},\delta_{z})$, where $E_{\rm GS}$ 
is the ground-state energy of the XXZ chain~(\ref{eq:XXZ}) per site 
and $E_{\rm GS}(\delta_{xy},\delta_{z})$ is that of the bond-alternating 
XXZ chain~(\ref{eq:dimer_XXZ}). If the dimerization parameter is small 
enough $|\delta_{xy,z}|\ll 1$, $\Delta E_{\rm GS}$ is expected to 
agree well with $\Delta {\cal E}_{\rm GS}$ in Eq.~(\ref{eq:E_GS}). 
In this case, we can extract the values of $d_{xy,z}$ from the relation 
$\Delta E_{\rm GS}=\Delta {\cal E}_{\rm GS}$.

To extrapolate the thermodynamic-limit value of 
$E_{\rm GS}(\delta_{xy},\delta_{z})$, we use Aitken-Shanks method 
for the results of finite-size numerical diagonalization, and 
the method works well since the bond-alternating chains are gapful. 
On the other hand, $E_{\rm GS}$ includes a large finite-size 
correction, as shown in Sec.~\ref{subsec:TLLandED}. 
Therefore, instead of numerically-evaluated $E_{\rm GS}$, 
we use its exact value fixed by Bethe ansatz~\cite{Yang66}
\begin{align}
\frac{E_{\rm GS}}{J}=&\frac{1}{4}\cos\gamma
-\frac{1}{2}\sin^{2}\gamma\nonumber\\
&\;\times\int_{-\infty}^{\infty}\frac{{\rm d}\lambda}{\cosh(\pi\lambda)}
\frac{1}{\cosh(2\gamma\lambda)-\cos\gamma},
\label{eq:E_GS_XXZ}
\end{align}
where $\gamma\equiv\cos^{-1}\Delta_{z}$. 
At the limit of $\gamma\to 0$, 
we obtain the ground-state energy for the Heisenberg model,
\begin{align}
\left.\frac{E_{\rm GS}}{J}\right|_{\gamma\to 0}=
\frac{1}{4}-\ln 2. 
\end{align}

\begin{figure}
\begin{center}
\includegraphics[width=0.4\textwidth]{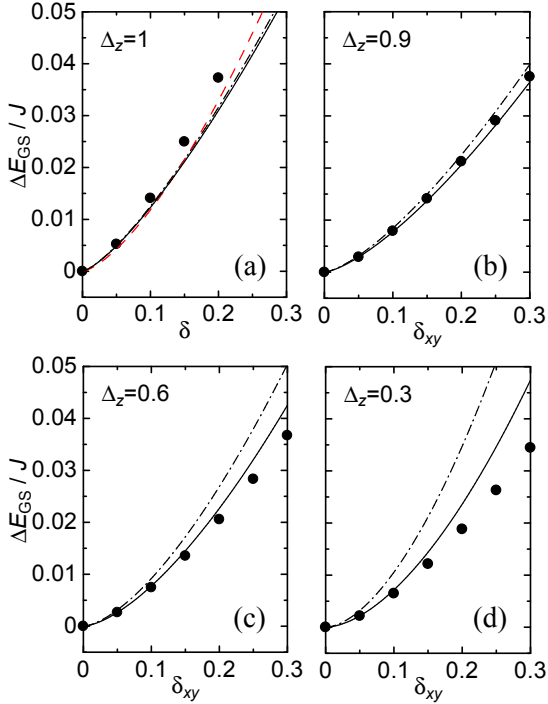}
\end{center}
\caption{(Color online) (a) Ground-state energy difference 
$\Delta E_{\rm GS}$ for the $SU(2)$-symmetric case with 
$\Delta_{z}=1$ and $\delta_{xy}=\delta_{z}=\delta$, 
obtained from numerical diagonalization (black circles). 
Solid and dashed-dotted curves represent Eq.~(\ref{eq:E_GS}) 
with $d_{xy[z]}$ determined from the relation 
$\Delta{\cal E}_{\rm GS}=\Delta E_{\rm GS}$ 
at $(\delta_{xy},\delta_z)=(0.05,0)$ [$=(0,0.05)$] 
and with those in Table~\ref{tb:dimer_coeff}, respectively. 
Dashed curve is Eq.~(\ref{eq:GS_SU2_dimer_corr}) including 
the logarithmic correction.
In the panels (b), (c), and (d), black circles are $\Delta E_{\rm GS}$ 
for $\Delta_{z}=0.9$, $0.6$, and $0.3$, respectively, under the 
condition $\delta_z=0$. Solid and dashed-dotted curves are respectively 
Eq.~(\ref{eq:E_GS}) with $d_{xy}$ obtained through 
$\Delta{\cal E}_{\rm GS}=\Delta E_{\rm GS}$ at $\delta_{xy}=0.05$ 
and with that in Table~\ref{tb:dimer_coeff}.} 
\label{fig:GSenergy}
\end{figure}

Black points in Fig.~\ref{fig:GSenergy} show $\Delta E_{\rm GS}$ 
determined from Eq.~(\ref{eq:E_GS_XXZ}) and numerically evaluated 
$E_{\rm GS}(\delta_{xy},\delta_z)$ for the cases of $\Delta_z=1$, 0.9, 0.6 
and $0.3$. The solid curve in the panel (a) of this figure represents 
the formula~(\ref{eq:E_GS}) with $d_{xy[z]}$ determined from 
$\Delta E_{\rm GS}$ at 
$(\delta_{xy},\delta_z)=(0.05,0)[=(0,0.05)]$. 
Solid curves in the panels (b), (c), (d) are also 
the formula~(\ref{eq:E_GS}) with $d_{xy}$ obtained in the same way. 
For comparison, we also draw dashed-dotted curves 
of the formula~(\ref{eq:E_GS}) with the coefficients in 
Table~\ref{tb:dimer_coeff}. In the $SU(2)$ case of the panel (a), 
the ground-state energy formula with a logarithmic correction 
\begin{align}
\frac{\Delta E_{\rm GS}}{J}=
\frac{0.2728\delta^{4/3}}{1+0.147\ln\left|\frac{0.1616}{\delta}\right|},
\label{eq:GS_SU2_dimer_corr}
\end{align}
which is predicted in Ref.~\onlinecite{Orignac04},
is also plotted as a dashed curve. 
As pointed out in Ref.~\onlinecite{Orignac04}, we find that even the curve 
including the correction deviates from the numerical data for 
$\delta\agt0.1$. 
On top of this isotropic case, 
Fig.~\ref{fig:GSenergy} shows that the accuracy of the fitting curves 
becomes worse as the anisotropy $\Delta_z$ decreases. This is a natural 
result from the fact that the formula~(\ref{eq:E_GS}) is broken down 
at the XY point with $\Delta_z=0$ and $\eta=1/2$. 
The deviation between the numerical data and the curve also becomes 
larger for $\delta\agt 0.1$ in the easy-plane region 
except for the case around $\Delta_z=0.9$. 
This sharply contrasts with the firm correspondence between 
dimerization gap and the sine-Gordon gap 
formula~(\ref{eq:dimergap}) (see, e.g, 
Figs.~\ref{fig:gap_XY}-\ref{fig:gap_Heisen_dimer}). 
We therefore determine the coefficients $d_{xy[z]}$ by using 
the numerical data $\Delta E_{\rm GS}$ for small 
dimerization parameters $(\delta_{xy},\delta_{z})=(0.05,0)[=(0,0.05)]$ 
or $(\delta_{xy},\delta_{z})=(0.1,0)[=(0,0.1)]$. 
They are summarized in Table~\ref{tb:dimer_coeff_GS}. 
There exists a large difference between $d_{xy,z}$ 
in Tables~\ref{tb:dimer_coeff} and \ref{tb:dimer_coeff_GS}, 
especially, in strongly easy-plane region.

In the remaining part of this subsection, we discuss the reason 
why $\Delta E_{\rm GS}$ fairly deviate from the analytic prediction 
$\Delta {\cal E}_{\rm GS}$ in contrast to the case of the dimerization 
gap in Secs.~\ref{subsec:XY_dimer}-~\ref{subsec:SU2_dimer}. 
Firstly, the sine-Gordon theory is just a perturbative 
low-energy effective theory for dimerized spin chains, while 
$\Delta E_{\rm GS}$ would be subject to 
high-energy states as well as low-energy ones. 
Therefore, it is expected that the formula~(\ref{eq:E_GS}) can be 
applicable only in an extremely weak dimerization regime. 
In fact, we find from Fig.~\ref{fig:GSenergy} that 
solid and dashed-dotted curves seem to become close to each other 
in an extremely weak dimerization regime $\delta_{xy},\delta\alt 0.05$. 
Hence, we conclude that it is dangerous to apply 
the sine-Gordon formula of the ground-state energy to 
the original spin chains with moderate dimerization. 
Secondly, the ground-state energy difference $\Delta E_{\rm GS}$ is always a 
convex-downward function of $\delta_{xy,z}$ in the whole region 
$0<\Delta_z\leq 1$. This convex property generally makes the accuracy 
of fitting decrease as the case of the dimerization gap 
in the ferromagnetic region $\Delta_z<0$. 
Moreover, as mentioned above, 
the formula~(\ref{eq:E_GS}) becomes invalid 
in the vicinity of both $\Delta_z=1$ and $\Delta_z=0$. 
From these arguments, coefficients $d_{xy,z}$ and $a_1$ obtained from 
low-lying excitation gaps are more reliable. 

\begin{table}[h]
\caption{\label{tb:dimer_coeff_GS} Dimer coefficients $d_{xy,z}$ 
of spin-$\frac{1}{2}$ XXZ chain obtained from ground-state energy 
difference $\Delta E_{\rm GS}$. 
The data outside (inside) the parentheses 
are fixed by the energy at $\delta_{xy,z}=0.05$ (0.1). 
The final line is the result for the $J$-$J_2$ model.}
\begin{ruledtabular}
\begin{tabular}{llll}
\multicolumn{1}{c}{$\Delta_z$} & \multicolumn{1}{c}{$d_{xy}$} & 
\multicolumn{1}{c}{$d_z$} & \multicolumn{1}{c}{$E_{\rm GS}-E_{\rm GS}(\delta_{xy},\delta_{z})$} \\
\hline
$1$       & 0.226 (0.239) & 0.107 (0.113) & $1.148(\delta_{xy}d_{xy}+\delta_{z}d_{z})^{1.333}$\\
$0.9$     & 0.261 (0.265) & 0.131 (0.134) & $1.331(\delta_{xy}d_{xy}+\delta_{z}d_{z})^{1.412}$\\
$0.8$     & 0.275 (0.274) & 0.143 (0.144) & $1.484(\delta_{xy}d_{xy}+\delta_{z}d_{z})^{1.459}$\\
$0.7$     & 0.283 (0.278) & 0.152 (0.151) & $1.673(\delta_{xy}d_{xy}+\delta_{z}d_{z})^{1.503}$\\
$0.6$     & 0.285 (0.278) & 0.159 (0.156) & $1.924(\delta_{xy}d_{xy}+\delta_{z}d_{z})^{1.550}$\\
$0.5$     & 0.284 (0.273) & 0.162 (0.158) & $2.280(\delta_{xy}d_{xy}+\delta_{z}d_{z})^{1.6}$\\
$0.4$     & 0.276 (0.264) & 0.163 (0.157) & $2.827(\delta_{xy}d_{xy}+\delta_{z}d_{z})^{1.656}$\\
$0.3$     & 0.262 (0.248) & 0.159 (0.132) & $3.766(\delta_{xy}d_{xy}+\delta_{z}d_{z})^{1.720}$\\
$0.2$     & 0.236 (0.221) & 0.149 (0.140) & $5.705(\delta_{xy}d_{xy}+\delta_{z}d_{z})^{1.796}$\\
$0.1$     & 0.188 (0.174) & 0.123 (0.114) & $11.72(\delta_{xy}d_{xy}+\delta_{z}d_{z})^{1.887}$\\
$0$       & $-$           & $-$           & $-$\\
$J$-$J_{2}$ & 0.342 (0.334) & 0.173 (0.171) & $1.265(\delta_{xy}d_{xy}+\delta_{z}d_{z})^{1.333}$
\end{tabular}
\end{ruledtabular}
\end{table}

\section{Applications}
\label{sec:apply}
In this section, we apply the results of Sec.~\ref{sec:delta} to 
some magnetic systems. We demonstrate that several physical 
quantities related to spins or dimerizations can be calculated 
accurately from the data in Tables~\ref{tb:dimer_coeff} and \ref{tb:a1}.

\subsection{Dimerized spin chains in a uniform field}
\begin{figure}
\begin{center}
\includegraphics[width=0.4\textwidth]{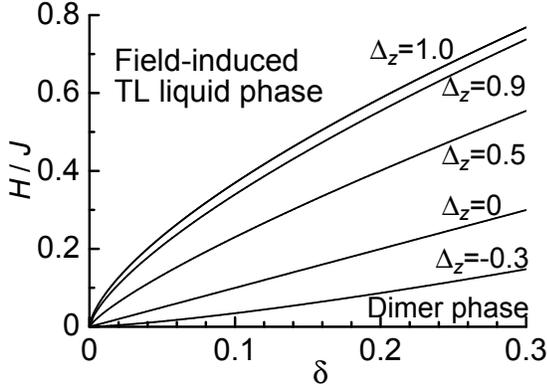}
\end{center}
\caption{(Color online) Ground-state phase diagram of 
the dimerized spin chains under a magnetic field $H$, 
Eq.~(\ref{eq:XXZ_dimer_field}). Each curve represents the phase boundary 
between the dimer and field-induced TLL phases.}
\label{fig:dimer_field}
\end{figure}

We first consider a spin-$\frac{1}{2}$ dimerized XXZ chain 
in a magnetic field. The Hamiltonian is defined as 
\begin{equation}
{\cal H}^{\delta\mathchar`-H} = 
{\cal H}^{{\rm XXZ}\mathchar`-\delta}-H\sum_j S_j^z,
\label{eq:XXZ_dimer_field}
\end{equation}
with $\delta_{xy}=\delta$ and $\delta_{z}=\Delta_{z}\delta$. 
As we have already explained, a spin gap opens in the zero-field case. 
However, a magnetic field $H>0$ 
induces the Zeeman splitting, and the gap of the magnon excitation 
with $S^z=1$ ($-1$) decreases (increases) as 
$\Delta E_{\Delta S_{\rm tot}^z=\pm1}\mp H$. 
When $H$ becomes larger than the value of the zero-field spin gap, 
the $S^z=1$ magnon condensation takes place and 
a field-induced TLL phase emerges with an incommensurate Fermi wave number 
$k_F=\pi-2\pi \langle S_j^z\rangle$. 
Therefore, the curve of the spin gap 
as a function of dimerization $\delta$ 
is directly interpreted as the ground-state phase boundary of 
the model~(\ref{eq:XXZ_dimer_field}), if the vertical axis (spin gap) is 
replaced with the strength of the magnetic field $H$. 
It is shown in Fig.~\ref{fig:dimer_field}. 

The critical point between the dimerized and TLL phases 
can be determined from experiments with varying $H$. 
Comparing the experimentally obtained critical field $H_c$ and 
the phase diagram of Fig.~\ref{fig:dimer_field} in 
quasi 1D dimerized spin-$\frac{1}{2}$ compounds, 
one can evaluate the strength of the dimerization $\delta$.

\subsection{Two-leg spin ladder with a four-spin interaction}
We next consider an $SU(2)$-symmetric two-leg spin-$\frac{1}{2}$ AF 
ladder with a four-spin exchange, whose Hamiltonian is given by 
\begin{align}
{\cal H}^{\rm lad} =& \sum_{j}\sum_{r=1,2}
J\bol{S}_{r,j}\cdot\bol{S}_{r,j+1} 
+\sum_j J_\perp \bol{S}_{1,j}\cdot\bol{S}_{2,j}\nonumber\\
&+\sum_j
J_4 (\bol{S}_{1,j}\cdot\bol{S}_{1,j+1})
(\bol{S}_{2,j}\cdot\bol{S}_{2,j+1}). 
\label{eq:ladder}
\end{align}
The symbol $r$ denotes the chain index. Three quantities $J>0$, $J_\perp$ 
and $J_4$ respectively stand for the intrachain-, interchain- and 
four-spin coupling constants. There are at least two kinds of 
physical origin of the four-spin term $J_4$. The first is that 
optical phonon modes with a spin-Peierls type coupling can cause a negative 
$J_{4}$.~\cite{Nersesyan97} 
The second is that the higher-order expansion of hopping terms 
in half-filled electron ladders 
with a strong on-site Coulomb repulsion.~\cite{Takahashi77,MacDonald88} 
In fact, the cyclic exchange term defined on each plaquette in the ladder 
contains a positive $J_4$ term, which is known to have scaling dimension 1 
and be most relevant in all the four-spin couplings of 
the cyclic term in the weak rung-coupling regime $J\gg |J_\perp|,|J_4|$.

The model~(\ref{eq:ladder}) has been analyzed 
by some groups.~\cite{Nersesyan97,Kolezhuk98,Nomura09} 
There appear four kinds of competing phases: the rung-singlet, Haldane, 
columnar-dimer, and staggered dimer phases.~\cite{Starykh04,Starykh07} 
In particular, the ground-state phase diagram in the region of 
$J_\perp>0$ and $J_4>0$ has been numerically 
completed in Ref.~\onlinecite{Nomura09}.

\begin{figure}
\begin{center}
\includegraphics[width=0.4\textwidth]{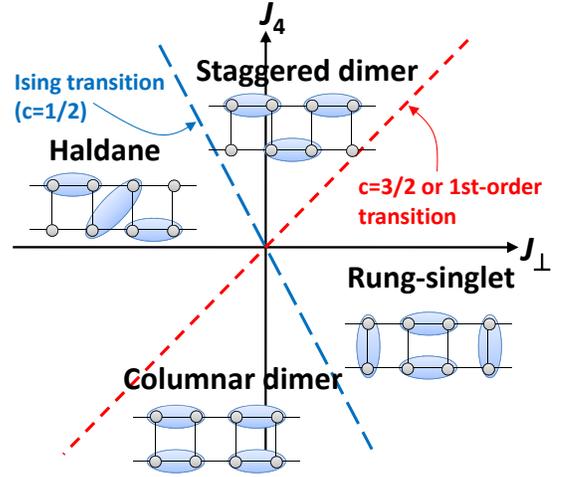}
\end{center}
\caption{(Color online) Ground-state phase diagram of the spin ladder~(\ref{eq:ladder}) 
in the weak rung-coupling regime. There are two transition curves, 
$J_4\approx 2.05 J_\perp$ and $J_4\approx -6.15 J_\perp$. 
The former is $c=3/2$ or first-order type, 
while the latter is in the Ising universality class with 
$c=1/2$ (see the text).}
\label{fig:ladder}
\end{figure}

Here, we show that the data in Tables~\ref{tb:dimer_coeff} and \ref{tb:a1} 
allow us to construct the phase diagram of the model~(\ref{eq:ladder}) 
in the weak rung-coupling regime with reasonable accuracy. 
From the bosonization, the low-energy effective Hamiltonian of 
Eq.~(\ref{eq:ladder}) reads
\begin{align}
{\cal H}^{\rm lad}_{\rm eff} =& \int dx \sum_{q=\pm}
\frac{v}{2}[K^{-1}(\partial_x\phi_q)^2+K(\partial_x\theta_q)^2]\nonumber\\
& +\frac{1}{a}(J_\perp\frac{\bar a^2}{2}-J_4\frac{(3d)^2}{2})
\cos(\sqrt{8\pi}\phi_+)\nonumber\\
&+\frac{1}{a}(J_\perp\frac{\bar a^2}{2}+J_4\frac{(3d)^2}{2})
\cos(\sqrt{8\pi}\phi_-)\nonumber\\
&+\frac{1}{a}J_\perp\bar a^2 \cos(\sqrt{2\pi}\theta_-)+\cdots. 
\label{eq:ladder_eff}
\end{align}
Here we have defined boson fields 
$\phi_\pm=(\phi_1\pm\phi_2)/\sqrt{2}$ and 
$\theta_\pm=(\theta_1\pm\theta_2)/\sqrt{2}$, where $\phi_r$ and $\theta_r$ 
are dual fields of the $r$-th chain (see Sec.~\ref{subsec:XXZ}). 
In Eq. (\ref{eq:ladder_eff}), we have extracted only the most relevant part 
of the rung couplings. The $SU(2)$ symmetry requires the relations 
$v=\pi Ja/2$, $K=1/2$, $a_1=b_0\equiv\bar a$ and $d_{xy}=2d_z\equiv 2d$. 
Due to this symmetry, three vertex terms in Eq.~(\ref{eq:ladder_eff}) 
have the same scaling dimension 1. 
The $(\phi_+,\theta_+)$ sector is equivalent to a sine-Gordon model. 
A Gaussian-type transition is expected at 
$J_\perp \bar a^2- J_4 (3d)^2=0$ if other irrelevant perturbations are 
negligible. On the other hand, the $(\phi_-,\theta_-)$ sector is 
a self-dual sine-Gordon model,~\cite{Lecheminant02} 
which is known to yield an Ising-type transition due to the competition 
between $\cos(\sqrt{8\pi}\phi_-)$ and $\cos(\sqrt{2\pi}\theta_-)$. 
The transition occurs as the strength of 
two coupling constants becomes equal, namely, 
$|J_\perp\bar a^2+ J_4 (3d)^2|/2=|J_\perp\bar a^2|$. 
Since we have already obtained the values of $\bar a$ and $d$ 
(see Tables~\ref{tb:dimer_coeff} and \ref{tb:a1}), 
we can draw the phase transition curves in 
the $J_\perp$-$J_4$ space in the weak rung-coupling regime, 
which are shown in Fig.~\ref{fig:ladder}. 
The two transition curves are represented as 
\begin{subequations}
\label{eq:curves}
\begin{align}
\label{eq:curve1}
J_4 =& \Big(\frac{\bar a}{3d}\Big)^2 J_\perp \approx 2.05 J_\perp,\\ 
J_4 =& -3 \Big(\frac{\bar a}{3d}\Big)^2 J_\perp \approx -6.15 J_\perp.
\label{eq:curve2}
\end{align}
\end{subequations}
Each phase is characterized by the locked boson fields and their position: 
In the columnar [staggered] dimer phase, 
$\phi_+$ and $\phi_-$ are respectively pinned at $\sqrt{\pi/8}$ and 0 
[$0$ and $\sqrt{\pi/8}$] and  
$(-1)^j\langle{\bol S}_{1,j}\cdot{\bol S}_{1,j+1}
+{\bol S}_{2,j}\cdot{\bol S}_{2,j+1}\rangle\propto 
\langle\sin(\sqrt{2\pi}\phi_+)\cos(\sqrt{2\pi}\phi_-)\rangle \neq0$ 
[$(-1)^j\langle{\bol S}_{1,j}\cdot{\bol S}_{1,j+1}
-{\bol S}_{2,j}\cdot{\bol S}_{2,j+1}\rangle\propto 
\langle\cos(\sqrt{2\pi}\phi_+)\sin(\sqrt{2\pi}\phi_-)\rangle \neq0$]. 
In the rung-singlet (Haldane) phase, $\theta_-$ is pinned instead of 
$\phi_-$ and $\langle\phi_+\rangle=\sqrt{\pi/8}$ ($0$), 
which corresponds to a non-zero ``even"-(``odd"-)type  nonlocal string 
order parameter.~\cite{Shelton96,Kim2000,Nakamura03}

It has been shown in Ref.~\onlinecite{Shelton96} that 
Eq.~(\ref{eq:ladder_eff}) can be fermionized. 
The resulting Hamiltonian consists of
three copies of massive Majorana fermions and another one 
(For detail, see e.g. Refs.~\onlinecite{Shelton96,Tsvelik,Gogolin}). 
The mass of the Majorana triplet $M_t$ and that of the remaining one 
$M_s$ are given by
\begin{subequations}
\label{eq:gap_ladder}
\begin{align}
M_t\propto& J_\perp\bar a^2- J_4 (3d)^2, \\
M_s\propto& 3 J_\perp\bar a^2+ J_4 (3d)^2. 
\end{align}
\end{subequations}
The transition curves in Fig.~\ref{fig:ladder} are identified 
with $M_t=0$ and $M_s=0$. 
At $M_s=0$, the low-energy physics is governed by the gapless 
singlet fermion which is equivalent to a critical Ising chain 
in a transverse field. The transition at $M_s=0$ therefore belongs to 
the Ising universality class with central charge $c=1/2$. 
On the other hand, three copies of massless Majorana fermions, 
which appear at $M_t=0$, are equivalent to an $SU(2)_2$ 
Wess-Zumino-Witten (WZW) theory~\cite{Tsvelik,Gogolin,Francesco} 
with central charge $c=3/2$. Thus, the transition at $M_t=0$ is expected 
to be a $c=3/2$ (first-order) type if the marginal current-current 
interaction~\cite{Shelton96,Starykh04,Starykh07}
omitted in Eq.~(\ref{eq:ladder_eff}) is 
irrelevant (relevant). In Ref.~\onlinecite{Nomura09}, 
the transition has been proved to be described by 
a $SU(2)_2$ WZW theory at least in the region of $J\gg J_\perp,J_4>0$. 
This suggests that the marginal term is irrelevant there. 
The Majorana fermion with the mass $M_t$ corresponds to 
a spin-triplet excitation (magnon), and 
another fermion with mass $M_s$ is a spin-singlet excitation, 
which is believed to be continuously connected to two-magnon bound 
state observed in the strong rung-coupling regime. 

Finally, we note that 
in the extremely weak rung-coupling limit, the coupling constants of 
vertex operators in Eq.~(\ref{eq:ladder_eff}) would be less valid 
since coefficients $\bar a$ and $d$ are determined from gaps 
induced by relatively large staggered field ($h_s/J=0.1$ or $0.3$) 
and dimerization ($\delta_{xy,z}=0.1$ or $0.3$), respectively. 
The true transition curves might somewhat deviate from our 
prediction~(\ref{eq:curves}). 
Our result is expected to be more reliable 
in a moderate rung-coupling regime. In fact, 
a numerical study in Ref.~\onlinecite{Nomura09} has shown that 
the phase boundary is located at $J_4/J_\perp\sim 2$ around $J_\perp/J=0.25$ 
(see Fig.~6 in Ref.~\onlinecite{Nomura09}), being consistent 
with Eq.~(\ref{eq:curve1}). 
We stress that our coefficients $\bar a$ and $d$ provides an easy way of 
estimating the phase boundary although it is a rough approximation compared 
with other sophisticated strategies such as DMRG and renormalization-group 
calculations. If we replace the intrachain term in Eq.~(\ref{eq:ladder}) 
with two $J$-$J_2$ chains~(\ref{eq:J-J2chain}), the intrachain 
marginal interaction omitted in Eq.~(\ref{eq:ladder_eff}) disappears. 
In this case, the prediction from the effective 
theory~(\ref{eq:ladder_eff}) becomes more reliable even in the weak 
rung-coupling limit $J_\perp/J,J_4/J\to0$. 
From the data of the $J$-$J_2$ model in 
Tables~\ref{tb:dimer_coeff} and \ref{tb:a1}, 
two transition curves in the modified ladder are 
\begin{subequations}
\label{eq:curves_v2}
\begin{align}
J_4 \approx& 0.69 J_\perp,\\ 
J_4 \approx& -2.08 J_\perp.
\end{align}
\end{subequations}

\subsection{Optical response of dimerized spin chains}
Optical responses in Mott insulators including multiferroic compounds 
have been investigated intensively. 
Quite recently, the authors in Ref.~\onlinecite{Katsura09} have 
theoretically studied the optical conductivity in a 1D 
ionic-Hubbard type Mott insulator with Peierls instability, 
whose strong coupling limit is equal to a 
spin-$\frac{1}{2}$ dimerized Heisenberg chain, ${\cal H}^{{\rm XXX}-\delta}$. 
The results in Ref.~\onlinecite{Katsura09} would be relevant to, 
for example, organic Mott insulators such as TTF-BA.~\cite{Kagawa10} 
In this system, the uniform electric polarization $P$ along the 1D chain 
is shown to be proportional to the dimer operator:
\begin{equation}
P = g a\sum_j(-1)^j {\bol S}_{j}\cdot{\bol S}_{j+1},
\end{equation}
where $g$ is the coupling constant between the polarization and dimer 
operators. Therefore, $P$ can be bosonized as 
\begin{equation}
P\approx 3d g \int dx \sin(\sqrt{4\pi}\phi(x))+\cdots,
\label{eq:P_bosonized}
\end{equation}
with $d_{xy}=2d$ and $d_{z}=d$. 
From Eq.~(\ref{eq:P_bosonized}), one can calculate $P$ and 
related observables by means of the bosonization 
for the dimerized spin chain. It has been shown that 
the spin-singlet excitation, i.e., the breather with mass $E_{B_2}$, 
is observed as the lowest-frequency sharp peak 
in the optical conductivity measurements. Since the mass $E_{B_2}$ is evaluated 
from the sine-Gordon theory as 
\begin{equation}
E_{B_2}/J=\sqrt{3}E_S/J=2.924 \delta^{2/3},
\end{equation}
we can extract the value of $\delta$ from the peak position 
of the optical conductivity. The exact expectation value of 
vertex operators in the sine-Gordon model has been predicted in 
Ref.~\onlinecite{Lukyanov97}. According to it, 
the polarization density is calculated to be 
\begin{equation}
\langle P\rangle/L=({\cal A}/3)^{3/2}(E_S a/v)^{1/2}3dg,
\end{equation}
with ${\cal A}\approx3.041$ and $L$ being the chain length. 
This provides an experimental way of estimating the coupling constant 
$g$, which is usually difficult to determine 
in other multiferroic compounds.

\section{Conclusions}
\label{sec:con}
We have numerically evaluated coefficients of 
bosonized dimer and spin operators in spin-$\frac{1}{2}$ XXZ 
model~(\ref{eq:XXZ}) and $J$-$J_2$ 
model~(\ref{eq:J-J2chain}), 
by using the correspondence between the excitation gap of deformed models 
with dimerization (or with staggered Zeeman term) and the gap formula 
for the sine-Gordon theory. 
This is a new strategy 
relying on a solid relationship between the lattice models 
and their low-energy effective theories. 
Our numerical approach is relatively easy compared with 
another method based on DMRG, developed in 
Refs.~\onlinecite{Hikihara98,Hikihara04}, although the accuracy 
is expected to be better in the latter method.
The obtained coefficients are summarized in 
Tables~\ref{tb:dimer_coeff} and \ref{tb:a1} and Fig.~\ref{fig:dimer_coeff}. 
In addition to these coefficients, 
we have pointed out a dangerous nature of applying 
the correlation amplitude~(\ref{eq:a1_anal}) 
as coefficients of bosonized spin operators near the 
$SU(2)$-symmetric point $\Delta_z=1$ in Sec.~\ref{subsec:spin_op}. 
Furthermore, we have also used the formula for ground-state 
energy of sine-Gordon model to calculate the same dimer coefficients
in Sec.~\ref{subsec:gs}. We conclude that the excitation-gap 
formula~(\ref{eq:dimergap}) is more suitable than the ground-state 
energy formula~(\ref{eq:E_GS}) for determining coefficients 
of bosonized operators.

Physical quantities associated with dimer and spin operators 
can be evaluated accurately by utilizing the dimer and spin coefficients. 
As examples, we have determined ground-state phase diagrams of 
dimerized spin chains in a uniform field and 
a two-leg spin ladder with a four-spin interaction 
in Sec.~\ref{sec:apply}. In addition, we have 
shown how to estimate the electromagnetic coupling constant and 
the strength of the dimerization from the optical observables 
in a ferroelectric dimerized spin chain. 
These applications clearly indicate high potential of the data in 
Tables~\ref{tb:dimer_coeff} and \ref{tb:a1}.

An interesting future direction is to apply a similar method to 
other 1D systems including fermion and boson models.  
Our method in this paper can be applied to lattice systems which 
have a well-established low-energy effective theory, in principle.

\begin{acknowledgements}
The authors thank Kiyomi Okamoto, Masaki Oshikawa, and T\^oru Sakai 
for useful comments. 
S.T. and M.S. were supported by Grants-in-Aid for JSPS Fellows 
(Grant No.\ 09J08714) and for Scientific Research from MEXT 
(Grant No.\ 21740295 and No. 22014016), respectively. 
The program package, TITPACK version 2.0, developed by Hidetoshi Nishimori, 
was used in the numerical diagonalization in Sec.~\ref{sec:delta}. 
\end{acknowledgements}


\end{document}